\documentclass[apj]{emulateapj}
\usepackage{lscape}
\shorttitle{ISM Properties of LGRB Host Galaxies}
\shortauthors{Levesque et al.}
\slugcomment{DRAFT \today}

\begin{document}

\title{The Host Galaxies of Gamma-Ray Bursts I: ISM Properties of Ten Nearby Long-Duration GRB Hosts}
\author{Emily M. Levesque$^{1,2}$}
\email{emsque@ifa.hawaii.edu}
\author{Edo Berger\footnotemark[2]}
\author{Lisa J. Kewley\footnotemark[1]}
\author{Megan M. Bagley\footnotemark[3]}

\begin{abstract}
We present the first observations from a large-scale survey of nearby ($z < 1$) long-duration gamma-ray burst (LGRB) host galaxies, which consist of eight rest-frame optical spectra obtained at Keck and Magellan. Along with two host galaxy observations from the literature, we use optical emission line diagnostics to determine metallicities, ionization parameters, young stellar population ages, and star formation rates. We compare the LGRB host environments to a variety of local and intermediate-redshift galaxy populations, as well as the newest grid of stellar population synthesis and photoionization models generated with the Starburst99/Mappings codes. With these comparisons we investigate whether the GRB host galaxies are consistent with the properties of the general galaxy population, and therefore whether they may be used as reliable tracers of star formation. Despite the limitations inherent in our small sample, we find strong evidence that LGRB host galaxies generally have low-metallicity ISM environments out to $z \sim 1$. The ISM properties of our GRB hosts, including metallicity and ionization parameter, are significantly different from the general galaxy population and host galaxies of nearby broad-lined Type Ic supernovae. However, these properties show better agreement with a sample of nearby metal-poor galaxies.
\end{abstract}

\footnotetext[1]{Institute for Astronomy, University of Hawaii,
2680 Woodlawn Dr., Honolulu, HI 96822}
\footnotetext[2]{Smithsonian Astrophysical Observatory, 60 Garden St., MS-20, Cambridge, MA 02138}
\footnotetext[3]{Department of Physics and Astronomy, University of Wyoming, Laramie, WY 82071}

\section{Introduction}
\label{Sec-intro}
Long-duration gamma-ray bursts (LRGBs) are currently used as tools for probing star-formation in distant galaxies (e.g., Wijers et al.\ 1998, Fynbo et al.\ 2007, Chen et al.\ 2005, Berger et al.\ 2006, Prochaska et al.\ 2007, Savaglio et al.\ 2009). These phenomena (with a burst duration of $>$ 2 s), are associated with the core-collapse of rapidly-rotating massive stars known as {\it collapsars} (Woosley 1993). LGRBs have also been correlated with the production of broad-lined Type Ic supernovae (e.g. Galama et al.\ 1998; Stanek et al.\ 2003; Kawabata et al.\ 2003; Hjorth et al.\ 2003). The short lifetimes of these massive progenitors suggest that LGRBs occur in star-forming galaxies such as young starbursts, setting them apart from the host galaxies of short-duration GRBs (Berger 2009). The correlation of LGRB optical afterglow locations to the UV light of their hosts also supports the connection with star formation (Bloom et al.\ 2002). Recently, Fynbo et al.\ (2007) reviewed recent work examining LGRBs and their host galaxies, and concluded that at high redshifts ($z > 2$), GRB hosts may be unbiased tracers of star formation, citing the relatively high amounts of star formation relative to luminosity for high-redshift galaxies with similar morphologies to the LGRB hosts. Chary et al.\ (2007) also show that there is good agreement between the star formation rate inferred by GRBs at z $\gtrsim$ 4 and the extinction corrected rates estimated from Lyman break galaxies. Finally, Savaglio et al.\ (2009) examine star formation properties for a large sample of archival GRB host galaxy observations, and conclude that the hosts are comparable to normal star-forming galaxies in both the local and distant universe.

In recent years, however, several studies have uncovered a connection between some LGRBs and low-metallicity galaxies that could threaten their utility as unbiased tracers of star formation in the universe. Stanek et al.\ (2006) found that the metallicities of five nearby LGRB hosts were lower than their equally-luminous counterparts, placing them below the standard luminosity-metallicity relation for dwarf irregular galaxies (Richer \& McCall 1995). Kewley et al.\ (2007) also placed LGRB host galaxies below the standard luminosity-metallicity (L-Z) relation for dwarf irregular galaxies, in a region of the diagram that also includes metal-poor galaxies. Fruchter et al.\ (2006) compared host galaxies of LGRBs and core-collapse supernovae and found that the LGRB environments are significantly different than those of core-collapse supernovae out to $z \sim 1$. They note that LGRBs were found in fainter and more irregular galaxies (Wainwright et al.\ 2007) and occurred in the brightest regions of their hosts, which are associated with concentrated populations of young massive stars. Modjaz et al.\ (2008) also found that LGRB host galaxies had systematically lower metallicities that the host galaxies of nearby ($z < 0.14$) broad-lined Type Ic supernovae with no accompanying GRB. Finally, Kocevski et al.\ (2008) model the mass-metallicity relation for LGRB host galaxies and find that at $z \lesssim 1$ LGRBs appear to be biasied towards low metallicity, though they suggest that this bias should disappear at higher redshifts as a result of metallicity evolution in the earlier universe.

Stellar evolutionary theory supports the possibility that the progenitors of LGRBs might favor low-metallicity environments. Late-type massive stars such as Wolf-Rayet stars are considered the most likely progenitor candidates (e.g., Hirschi et al.\ 2005, Yoon et al.\ 2006, Langer \& Norman 2006, Woosley \& Heger 2006). Stellar winds in these stars are driven by radiation pressure on spectral lines, which in turn affect the mass loss rates (Vink \& de Koter 2005). As a result, a star's wind-driven mass loss is heavily dependent on its surface metallicity (Kudritzki 2002). Vink et al.\ (2001), who accommodate the important fact that terminal velocity is weakly dependent on metallicity ($v_{\infty} \propto Z^{0.13}$; Leitherer et al.\ 1992), determine a mass loss-metallicity relation of $\dot M_w \propto Z^{0.7}$. As a result, surface velocities are higher for Wolf-Rayet stars at low metallicities, a consequence of the lower mass loss rate and a potentially important property of collapsars (Kudritzki \& Puls 2000, Meynet \& Maeder 2005). These surface velocities in turn generate a rotation-driven mass loss component (Meynet \& Maeder 2000). Some mass loss in LGRB progenitors is required - the LGRB-associated supernovae that have been identified are all categorized as broad-lined Type Ic supernovae, and production of these supernovae implies that the progenitors must undergo a level of mass loss sufficient to shed their H and He envelopes prior to core-collapse.

Recent progress has been made in satisfying these complicated collapsar parameters. Models of late-type massive stars in low-metallicity environments have managed to sustain high rates of rotation while still shedding their hydrogen envelopes (Yoon, Langer, \& Norman 2006, Woosley \& Heger 2006). Red supergiants in low-metallicity galaxies are found to exhibit unique physical instabilities and episodes of high mass loss that can be associated with the evolutionary limitations of their environment (Levesque et al.\ 2007), offering observational evidence of the extreme effect that low-metallicity environments are expected to have on the later phases of stellar evolution (Leitherer 2008). Combined, these arguments present compelling evidence that a low-metallicity environment may help massive stars evolve into LGRB progenitors.

However, it is also possible that the observed connection between LGRBs and low-metallicity environments may not be a direct result of metallicity at all, but rather an artifact of other galactic properties that might be favored by LGRBs. It has been proposed that this apparent low-metallicity bias is instead an artifact of an age bias. LGRBs are expected to occur in galaxies with younger stellar populations (which typically have lower metallicities) as a possible result of their massive progenitors' short lifetimes (Bloom et al.\ 2002, Berger et al.\ 2007). LGRBs may also favor such galaxies because of their association with burst-like star formation histories, as bursts of star formation can generate dense clusters of massive stars. This explanation considers both the standard collapsar model of GRB progenitors and a second model that generates LGRBs during the formation of black hole X-ray binaries (van den Heuvel \& Yoon 2007).

Determining the nature of the relationship between LGRBs and low-metallicity host environments is extremely important. A large-scale metallicity bias extended to higher redshifts could challenge the use of these phenomena as tracers of star formation in normal galaxies at large look-back times. Such a result would suggest that LGRBs are not the best means of probing early star formation, since they would be considerably less likely to occur in normal star-forming galaxies (Stanek et al.\ 2006). Conversely, it is also possible that a low-metallicity bias would only present a problem for nearby LGRB hosts, since at redshifts of $z > 1$ the mean metallicity for galaxies is lower as a whole (e.g., Kobulnicky \& Kewley 2004, Shapley et al.\ 2004, Erb et al.\ 2006, Chary et al.\ 2007, Liu et al.\ 2008). A decrease in metallicity at larger redshifts suggests that a low-metallicity trend may not necessarily bias LGRBs as high-redshift star-formation tracers (Berger et al.\ 2007, Fynbo et al.\ 2006, Kocevski et al.\ 2009).

To determine whether LGRB hosts are representative of the general galaxy population requires a large number of uniform high-quality rest-frame optical spectra of LGRB host galaxies, in order to gain a comprehensive picture of their metallicity and other ISM properties. Here we present our first observations from a large-scale survey of nearby ($z < 1$) LGRB host galaxies, which consist of eight new LGRB host galaxy observations, as well as two spectra of nearby LGRB hosts from the literature. We utilize our new grid of stellar population synthesis and photoionization models that has recently been made available (Levesque et al.\ 2009), using the Starburst99 evolutionary synthesis code (Leitherer et al.\ 1999, V\`{a}zquez \& Leitherer 2005) and the Mappings III shock and photoionization code (Binette et al.\ 1985, Sutherland \& Dopita 1993). This grid spans ages from 0 to 10 Myr, accommodating the age range of the galaxies' youngest massive star populations, and adopts a rich variety of ISM properties and star formation histories in order to accurately model a wide range of galaxy populations. This is the first grid of photoionization models that shows agreement with the observed emission line properties of low-metallicity galaxies, since previous models did not generate a sufficiently hard ionizing radiation spectrum to reproduce the required fluxes (Kewley et al.\ 2001, Levesque et al.\ 2009). With such models in place, the ISM properties of LGRB host galaxies, as well as the environments of galaxies from the general population, can be compared in detail.

We detail our sample of LGRB host galaxies and comparison galaxy populations from the literature (\S~2) and describe how we determine the ISM properties for these galaxies (\S~3). We then compare the LGRB hosts to our various galaxy samples, examining their ISM properties and placing them on a variety of emission line diagnostic diagrams (\S~4). We compare the LGRB host galaxies to Starburst99/Mappings grids (\S~5). Finally, we discuss the implications of our results and future work in this area (\S~6). Throughout this work we assume a cosmology of $H_0 = 70$ km s$^{-1}$ Mpc$^{-1}$, $\Omega_m = 0.3$, and $\Omega_\Lambda = 0.7$.

\section{GRB Host and Comparison Samples}
\label{data}
\subsection{Nearby LGRB Host Galaxy Survey}
We are conducting a uniform rest-frame optical spectroscopic survey of nearby LGRB host galaxies using the Keck telescopes on Mauna Kea and the Magellan telescopes at Las Campanas Observatory. The sample included in this survey was compiled from the GHostS database (Savaglio et al.\ 2006) and the Gamma-Ray Burst Coordinates Network and restricted to confirmed host galaxies of long-duration ($>$ 2 s) GRBs with redshifts of $z < 1$, allowing us to obtain rest-frame optical spectra from 3000-7000\AA\ using optical and near-infrared observations. Eight galaxies from this ongoing survey are included here.

\subsubsection{Keck: GRBs 991208, 010921, 020903, 031203, 030329, 051022, and 060218}
Seven LGRB host galaxy spectra were obtained using the Low-Resolution Imaging Spectrograph (LRIS) on the Keck I telescope at Mauna Kea. We used the long 1" slitmask for our observations. To calibrate the observations we obtained internal flat fields and comparison lamp spectra with the standard Hg, Ne, Ar, Cd, and Zn lamp setup available at LRIS. We observed spectrophotometric standards to flux-calibrate the host spectra. The dates and details of our observations are given in Table \ref{tab:params}. In most of these cases, the host galaxies were quite dim ($V \sim$ 21 to 25 mag). In order to ensure that we successfully acquired these host galaxies in the slit, we first centered on a nearby bright star. From this star we calculated the distance to the GRB host and the position angle that would place both the bright star and the host on the slit, ensuring that we observed spectra of both objects. As a result, we did not observe at the parallactic angle.

\subsubsection{Magellan: GRB 020405}
A spectrum of GRB 020405 was obtained with LDSS3 mounted on the Magellan/Clay 6.5m telescope on 2008 January 16 UT with a total exposure time of 4500 s.  We used the VPH-Red grism with an OG590 order blocking filter and a 1'' slit.  The data were reduced using standard IRAF packages, and wavelength calibration was performed using HeNeAr arc lamps.  The resulting spectrum covers 6200-8600 \AA.

\subsubsection{Published LGRB Host Spectra: GRB 980425 and GRB 997012}
Emission line fluxes for GRB 980425 and GRB 997012 were taken from previous observations published in Christensen et al.\ (2008), and K\"{u}pc\"{u} Yoldas et al.\ (2006) respectively. Christensen et al.\ (2008) obtained integral field spectroscopy of the host of GRB 980425 in April and May 2006 using VIMOS at the VLT; we adopt their published fluxes corrected for extinction. K\"{u}pc\"{u} Yoldas et al.\ (2006) obtained spectra of the host of GRB 997012 $\sim$6 years after the burst, scrutinizing how the fluxes of several strong emission lines varied with time. We adopt emission line fuxes from their spectrum observed on 2005 July 5-6 with FORS2 at the VLT.

\subsubsection{Data Reduction}
We reduced and analyzed the LGRB host galaxy data from Keck and Magellan using IRAF\footnotemark[1]. \footnotetext[1]{IRAF is distributed by NOAO, which is operated by AURA, Inc., under cooperative agreement with the NSF.} We used the \texttt{lrisbias} IRAF task distributed by the W. M. Keck Observatories to subtract overscan from the LRIS images, and applied a flatfield correction based on the internal lamp flats. The spectra were extracted using an optimal extraction algorithm, with deviant pixels identified and rejected based upon the assumption of a smoothly varying profile. For the dimmest host galaxies, we used the spectrum of a nearby bright star placed on the slit during observations to determine an accurate trace for extraction. Wavelength calibration was performed based on our comparison lamp observations, and flux calibration was based on our observations of spectrophotometric standards. In most cases, emission line fluxes were determined using the IRAF task \texttt{splot} in the \texttt{kpnoslit} package to fit Gaussians to the line profiles. In cases where the emission lines were found to have asymmetric shapes (GRB 991208, GRB 030329, GRB 060218; see the Appendix for discussion), the fluxes were determined by fitting the lines with the sum of multiple Gaussians, using the IRAF task \texttt{ngaussfit} in the \texttt{stsdas.analysis.fitting} package. The raw fluxes that we measure for each of our observed host galaxies are given in Table \ref{tab:fluxes}.

\subsection{Comparison Samples}
We compare the ISM properties of our LGRB host galaxies to a variety of other galaxy populations. For all of these populations, we restricted our samples to star-forming galaxies using the emission-line diagnostics of Kewley et al.\ (2006). This classification is based on the empirical criteria of Kauffmann et al.\ (2003) derived from the [NII]/H$\alpha$ vs. [OIII]/H$\beta$ diagnostic diagram, and the theoretical Kewley et al.\ (2001) criteria based on the [SII]/H$\alpha$ vs. [OIII]/H$\beta$ and [OI]/H$\alpha$ vs. [OIII]/H$\beta$ diagnostic diagrams. Equations quantifying these criteria are given in Kewley et al.\ (2006). This classification removes contamination from composite galaxies, Seyferts, LINERs, and ambiguous galaxies that cannot be definitively classified using optical diagnostics alone.

{\it Sloan Digital Sky Survey (SDSS) and Nearby Field Galaxy Survey (NFGS) galaxies}: To compare our LGRB hosts to the general local ($z \le 0.1$) star-forming galaxy population, we plot emission line ratios determined for a sample of emission-line galaxies from SDSS described in Kewley et al.\ (2006). These galaxies were originally taken from Data Release 4 of SDSS (Adelman-McCarthy et al.\ 2006) and restricted to  85224 galaxies with a S/N $\ge$ 8 in the strong emission lines and a redshift range between 0.04 $< z <$ 0.1. The lower limit of this redshift range corresponds to an aperture covering fraction of 20\%, the minimum required to avoid domination of the spectrum by aperture effects (Kewley et al.\ 2005, Kewley \& Ellison 2008). We further restrict this sample to the 60920 galaxies classified as star-forming by Kewley et al.\ (2006).

To ensure that the local galaxy population comparison is not affected by any residual aperture effects present in the SDSS sample, we include also a sample of galaxies from NFGS. NFGS has integrated emission-line spectrophotometry of 196 galaxies that span from $-14 < M_B < -22$ and include the full range of Hubble sequence morphologies, with a median redshift of $z = 0.01$ and a maximum redshift of $z \sim 0.07$ (Jansen et al.\ 2000a, 2000b). The galaxies were selected from the CfA redshift catalog, which has a limiting blue photographic magnitude of $m_Z = 14.5$ (Huchra et al.\ 1983) and observed with the FAST spectrograph at the Whipple 1.5 m telescope (Jansen et al.\ 2000b). We limit this sample to 95 star-forming galaxies based on the Kewley et al.\ (2006) criteria.

{\it Blue Compact Galaxies (BCGs)}: To compare the LGRB hosts to a sample of galaxies that show evidence of burst-like star formation histories, we include a sample of blue compact galaxies (BCGs) from Kong \& Cheng (2002) and Kong et al.\ (2002). Their sample consists of 97 galaxies selected from the catalogs of Gordon \& Gottesman (1981), Thuan \& Martin (1981), Kinney et al.\ (1993), and Thuan et al.\ (1999), limited to $M_B < -17$ mag. Here we restrict our sample to those galaxies with a complete set of spectral line fluxes that satisfy the star-forming galaxy emission line criteria in Kewley et al.\ (2006), giving us 36 star-forming BCGs in our final sample.  We note that, for consistency with our other star-forming galaxy samples, we use the Kewley et al.\ (2006) criteria rather than the classifications use in Kong et al.\ (2002) to remove AGN and non-emission-line galaxies from their sample. This sample of BCGs has a redshift range of $0.003 < z < 0.029$, with a median redshift of $z = 0.016$.

{\it Type Ic SN Host Galaxies (SNHGs)}: Modjaz et al.\ (2008) present a sample of nearby host galaxies where broad-lined Type Ic supernovae have been observed {\it without} accompanying GRBs. From their sample we select eight SNHGs with the full set of published emission line fluxes required for our analyses, including additional fluxes in cases where observations of the SN position differed from the galaxy center. To better clarify selection effects, Modjaz et al. (2008) distinguish between host galaxies of supernovae that were detected through galaxy-targeted searches (and therefore reflect a selection bias towards more luminous host galaxies), and supernovae detected in large-scale surveys, which have a slightly fainter mean luminosity and selection criteria that are more comparable to GRBs. Our subsample includes two hosts of targeted-search supernovae and six hosts of supernovae detected through surveys, and ranges from $-20.3 < M_B < -16.9$. The SNHGs used here have a redshift range of $0.012 < z < 0.137$, with a median redshift of $z = 0.058$.

{\it Metal-Poor Galaxies (MPGs)}: We include a sample of 10 MPG spectra from Brown et al.\ (2008). The 10 MPG spectra are part of a survey of 38 MPG candidates, selected from Data Release 4 of SDSS based on restrictions on their ($u' - g'$)$_0$, ($g' - r'$)$_0$, and ($r' - i'$)$_0$ colors, a magnitude limit of $g' < 20.5$, and removal of HII regions in nearby galaxies through visual inspection. The remaining MPG candidates were observed with the Blue Channel spectrograph at the MMT telescope. Brown et al.\ (2008) find log(O/H) + 12 $<$ 8 based on their electron temperature metallicities (Izotov et al.\ 2006). The MPG sample has a redshift range of $ 0.03 < z < 0.081$ with a median redshift of $z = 0.073$.

{\it Team Keck Redshift Survey (TKRS) galaxies}: For a comparison of our intermediate-redshift ($0.3 < z < 1$) LGRB host galaxies with similar star-forming galaxies, a deeper survey is required. The Great Observatories Origins Deep Survey-North (GOODS-N; Giavalisco et al.\ 2004) includes 0.3 $< z <$ 1 emission-line galaxies and is 90\% complete at $Z_{AB} = 24$. Kobulnicky \& Kewley (2004) measured nebular oxygen abundances for 204 of these galaxies using publicly available spectra from the TKRS observations using DEIMOS on Keck II (Wirth et al.\ 2004). The TKRS sample has a median redshift of $z = 0.65$, and is 53\% complete at its limiting magnitude of R$_{AB}$ = 24.4.

\section{Analysis of ISM Properties}
\label{ISM}
\subsection{Emission Line Flux Measurements}
Since we adopt the ``observed" fluxes for the SNHGs, MPGs, and BCGs from the literature, as well as our LGRB host sample, we correct all of these fluxes for local extinction effects using the observed Balmer lines and the Cardelli et al.\ (1989) reddening law with the standard total-to-selection extinction ratio $R_V = 3.1$. We first calculated E($B-V$) with the equation
\begin{equation}
E(B-V) = \frac{\rm{log}(\frac{C}{X/H\beta})}{0.4 \times (k(X) - 3.609)}
\end{equation}

where X is a Balmer line flux (H$\alpha$, H$\gamma$, or H$\delta$), C is the Balmer decrement of the
ratio X/H$\beta$ for case B recombination (H$\alpha$/H$\beta$ = 2.87, H$\gamma$/H$\beta$ = 0.466, and H$\delta$/H$\beta$ = 0.256 for T$_e = 10^4$ K and $n_e\sim 10^2$ - $10^4$ cm$^{-3}$, following Osterbrock 1989) and k(X) is the wavelength-dependent constant for X from Cardelli et al.\ (1989) (k(H$\alpha$) = 2.535, k(H$\gamma$) = 4.174, and k(H$\delta$) = 4.438). For some of our higher-redshift LGRB host galaxies H$\alpha$ was not observed; in those cases we use the H$\gamma$ flux to determine E($B-V$). In the case of GRB 020405, where there was no detection of H$\gamma$, we use the H$\delta$ flux. Once E($B-V$) was determined for a galaxy, the observed emission line fluxes were then dereddened using the \texttt{ccm\_unred} function in IDL. Values for E($B-V$) are given in Table \ref{tab:gals}.

\subsection{Metallicity}
We used the extinction-corrected emission line fluxes to determine the ISM properties for these galaxies listed in Table \ref{tab:gals}. We determine metallicities using the R$_{23}$ diagnostic put forth by Kewley \& Dopita (2002) and refined by Kobulnicky \& Kewley (2004). Since this diagnostic is double-valued we use several tests to determine whether each galaxy's metallicity should be determined by the ``upper" branch or ``lower" branch equations of the diagnostic. The presence of the auroral [OIII]$\lambda$4363 emission line is often a good indication that the galaxy has a low metallicity, as the weakness of the line renders it unobservable in higher-metallicity galaxies at the S/N and sensitivity of our spectra (Garnett et al.\ 2004). Where possible, the [NII]/[OII] ratios were used to differentiate between the upper and lower branches of the R$_{23}$ diagnostic, with log([NII]/[OII]) $> -1.2$ indicating upper branch and log([NII]/[OII]) $< -1.2$ indicating lower branch (Kewley \& Ellison 2008). The [NII]/H$\alpha$ ratio provided a third means of determining the diagnostic branch, with log([NII]/H$\alpha$) $> -1.1$ indicating upper branch and log([NII]/H$\alpha$) $< -1.3$ indicating lower branch (Kewley \& Ellison 2008), leaving an indeterminate range of values in between. The ionization parameter $q$ was determined using the Kewley \& Dopita (2002) [OIII]/[OII]-$q$ relation. Here we define $q$ in cm s$^{-1}$ as the maximum velocity possible for an ionization front being driven by the local radiation field, where $q$ relates to the dimensionless ionization parameter ($\mathcal{U}$) by $\mathcal{U} \equiv q/c$.

For comparison where possible, we also calculated metallicities using the Pettini \& Pagel (2004) relation between log([OIII]/H$\beta)$/(NII]/H$\alpha$) ($O3N2$) and metallicity, where 12 $+$ log(O/H) $ = 8.73 - 0.32 \times O3N2$. We applied this relation for all galaxies whose spectra included [NII] line fluxes. These metallicities are also included in Table \ref{tab:gals}.

Finally, for three of our LGRB hosts (GRBs 030329, 031203, and 060218) and the MPG sample, we detect the auroral [OIII] $\lambda$4363 emission line. The presence of [OIII]$\lambda$4363 allows us to calculate the electron temperature ($T_e$) metallicities for these three hosts. Higher chemical abundances increase the rate of nebular cooling in galaxies, lowering temperatures in HII regions. As a result, the oxygen abundance (and therefore an estimate of $T_e$) can be measured from the ratio of [OIII]$\lambda$4363 to lines with a lower excitation potential, such as [OIII]$\lambda$5007 and [OIII]$\lambda$4959. This yields $T_e$(O$^{++}$), which we used to calculate $T_e$(O$^{+}$) by the relation $T_e$(O$^{+}$) = 0.7$T_e$(O$^{+}$) $+ 0.3$ from Stasinska (1980).

The electron density $n_e$ was estimated from the ratio of the [SII]$\lambda$6717/[SII]$\lambda$6731 doublet lines. With these ratios, we determined values for $n_e$ and $T_e$ using the IRAF task \texttt{temden} in the \texttt{stsdas.analysis.nebular} package. In the case of GRB 030329, we do not detect the [SII] doublet; instead we calculate $T_e$ assuming electron densities of both 100 cm$^{-3}$ and 200 cm$^{-3}$ (in agreement with our other $n_e$ values) and find identical results, as $T_e$ is insensitive to small changes in $n_e$ (Kewley et al.\ 2007). These parameters were used in equations for the abundances of O$^{++}$/H$^+$ and O$^+$/H$^+$ (Shi, Kong, \& Cheng 2006, Garnett 1992). Once these abundances were determined the log(O/H) + 12 values could be calculated. It should be noted that this method is known to yield systematically lower metallicities than those determined from emission-line diagnostics (Kennicutt, Bresolin, \& Garnett 2003; Kewley \& Ellison 2008).

\subsection{Star Formation Rates}
At present, the (dereddened) H$\alpha$ 6563\AA\ emission feature is the most reliable optical tracer of the star formation rate (SFR) in a galaxy, because H$\alpha$ scales directly with the total ionizing flux of newly-formed stars in the ionization nebulae of star-forming galaxies and HII regions (e.g., Kennicutt 1998, Kewley et al.\ 2004). However, for galaxies with $z \gtrsim 0.3$ the H$\alpha$ line is redshifted out of the optical regime, making it impractical as a SFR indicator. A popular alternative for higher-redshift galaxies is the [OII] $\lambda\lambda$3727,3729 doublet. However, this line is strongly dependent on $T_e$, and hence the chemical abundance of the galaxy, requiring that accurate relations between [OII] and SFR take metallicity into account. Considering this dependence, Kewley et al.\ (2004) determined a metallicity-depedent SFR calibration for [OII].

We calculate H$\alpha$ SFRs using the relation of Kennicutt (1998) where SFR(M$_{\odot}$/yr) = (7.9 $\times$ 10$^{-42}$) $\times$ L(H$\alpha$). When the H$\alpha$ flux is not available, as was the case for many of our LGRB host galaxies (50\%), we instead use the metallicity-dependent SFR relation for the [OII]$\lambda$3727 luminosities from Kewley et al.\ (2004). The H$\alpha$ luminosities (when available) and SFRs are given in Table \ref{tab:gals}. For some of our LGRB host galaxies, we also determine H$\alpha$-based SFRs using the H$\beta$ emission line flux, assuming a Balmer decrement of 2.87 (Osterbrock 1989) to approximate the flux of the H$\alpha$ line; these SFRs are included in Appendix A. These H$\beta$-derived SFRs assume no contamination from an underlying old stellar population, consistent with a young stellar population.

\subsection{Young Stellar Population Ages}
We determined the young stellar population ages for the LGRB hosts, MPGs, and BCGs using their rest-frame values for the equivalent width of the H$\beta$ emission line ($W_{H\beta}$). Copetti et al.\ (1986) find that $W_{H\beta}$ decreases nearly monotonically with the age of an HII region. This parameter is largely independent of the electron temperature $T_e$ and electron density $n_e$, but sensitive to the initial mass function, stellar mass loss rate, and metallicity. However, $W_{H\beta}$ is primarily dependent on the evolution of the HII region, and as such is a strong indicator of the young stellar population age.

Schaerer \& Vacca (1998) use the Geneva HIGH evolutionary tracks in conjunction with a series of theoretical stellar spectra (Schaerer et al.\ 1996a, 1996b; Schaerer \& deKoter 1996; Kurucz 1991; Schmutz et al.\ 1992), and observed Wolf-Rayet emission lines to construct evolutionary synthesis models for young starburst galaxies. From these models they plot the evolution of $W_{H\beta}$ for the Geneva group metallicities (see Figure 7 of Schaerer \& Vacca 1998). Using these data, we derive equations that give a quantitative relation between $W_{H\beta}$ and young stellar population age for galaxies at each of the Geneva tracks' metallicities ($Z/Z_{\odot}$ = 0.05, 0.20, 0.40, 1.00, and 2.00), with the hopes that they will be useful tools for determining the young stellar population ages of starburst galaxies at various metallicities in the future. The equations take the form:

\begin{eqnarray}
\rm{log(Age)} = \rm{A} + \rm{B}W_{H\beta} + \rm{C}W_{H\beta}^2  + \rm{D}W_{H\beta}^3 + \rm{E}W_{H\beta}^4 \nonumber\\
\noindent + \rm{F}W_{H\beta}^5 + \rm{G}W_{H\beta}^6 + \rm{H}W_{H\beta}^7 + \rm{I}W_{H\beta}^8 + \rm{J}W_{H\beta}^9
 \pm \rm{K},
\end{eqnarray}

where the constants (A-J) and the error of the fit (K) are given in Table \ref{tab:ages} for the various metallicities. The high-order polynomial is necessary to preserve the fluctuating contributions of the Wolf-Rayet-rich stellar evolutionary phase between $\sim$ 2 - 7 Myr; additional discussion of stellar population contributions in an instantaneous burst star formation history is given in Levesque et al.\ (2009).

It is important, however, to note the shortcomings of such a method. The progression of $W_{H\beta}$ with age necessitates the assumption of a zero-age instantaneous burst star formation history (Copetti et al.\ 1986), since it decreases monotonically with the age of the burst due to a decrease in ionizing photons and an increase in regional continuum luminosity. While this assumption is robust across variations in the initial mass function (Stasinska \& Leitherer 1996), it also assumes complete absorption of the ionizing photons and uniform extinction across nebular and stellar emission, both of which are not necessarily true. 

\subsection{AGN Activity in the Host of GRB 031203}
During our emission line analyses, we found that the host galaxy of GRB 031203 displayed several unusual emission line ratios that were not consistent with the other LGRB host galaxies or the star-forming galaxies in our comparison samples. Instead, these ratios suggested that GRB 031203 might show evidence of AGN activity. To determine whether or not GRB 031203's host was a typical star-forming galaxy, we applied the emission line ratio classification scheme from Kewley et al.\ (2006) used to separate star-forming galaxies, composite galaxies, and Seyfert or Low Ionization Narrow Emission-Line Region (LINER) AGN. These diagnostics utilize the [OIII]$\lambda$5007/H$\beta$, [NII]$\lambda$6584/H$\alpha$, [SII]$\lambda\lambda$6717,6731/H$\alpha$, and [OI]$\lambda$6300/H$\alpha$ line ratios (GRB 031203 was the only galaxy in our LGRB host sample that included a detection of [OI] $\lambda$6300). Based on these diagnostics and our measured line fluxes, the host of GRB 031203 appears to show definitive evidence of AGN activity, and {\it cannot} be classified as a purely star-forming galaxy. The exact AGN classification of the host cannot be determined without further analysis, as even small uncertainties in the line fluxes can accommodate either a composite or Seyfert galaxy scenario (the host does not agree with the criteria for LINER galaxies). This is the first evidence of AGN activity in a LGRB host galaxy.

This is at odds with the results of Margutti et al.\ (2007) and Prochaska et al.\ (2004), who both classify the host of GRB 031203 as star-forming. However, these classifications fail to accommodate for the sometimes considerable error bars on the published emission line fluxes. Based on the Kewley et al.\ (2006) criteria, and taking the flux errors into consideration, the Prochaska et al.\ (2004) and the Margutti et al.\ (2007) fluxes cannot be used to distinguish whether the host of GRB 031203 is a star-forming, composite, or Seyfert galaxy.

While this result is intriguing, it must be noted that the emission-line diagnostics used to determine properties such as metallicity are meant for use with star-forming galaxies. We consider the ISM properties derived in this paper to be good {\it approximations} of the ISM environment of GRB 031203; however, we cannot quantify how the presence of AGN activity might affect these results. Therefore, in our comparisons with the general galaxy population we must be careful to consider how the inclusion or exclusion of GRB 031203's host in the LGRB host sample affects our overall conclusions.

\section{Results for LGRB Hosts and Comparison With Galaxy Samples}
\label{comps}
\subsection{ISM Properties}
\subsubsection{Luminosity vs. Metallicity}
In Figure 1 (top), we place our low-redshift ($z < 0.3$) LGRB hosts on a luminosity-metallicity (L-Z) plot, and compare their position to the BCGs, MPGs, and SNHGs included in our sample. All of the metallicities plotted in Figure 1 were calculated using the R$_{23}$ diagnostic calibration described in 3.2. We illustrate the L-Z relation for each population, and find that the BCGs and SNHGs follow a very similar L-Z relation. By contrast, the MPGs and LGRB host galaxies lie well below the L-Z relation of the BCGs, with the MPGs falling slightly lower than the LGRB hosts at the same $M_B$ on the diagram. Although these conclusions are certainly limited by the small LGRB host sample size, this is a similar result to that found in Kewley et al.\ (2007), where LGRB hosts and metal-poor galaxies are found to lie at systematically lower metallicites for their luminosity as compared to the L-Z relation for dwarf irregular galaxies. Our work shows no similar skew towards lower metallicities in the SNHGs. This is also in agreement with the findings of Modjaz et al.\ (2008), who conclude that the SNHGs have higher metal abundances than LGRB host galaxies and occupy a distinct and separate higher-metallicity region of the L-Z plot as compared to LGRB host galaxies. From an application of the two-sample Kolmogorov-Smirnov (KS) test using these galaxies' metallicities, we find that the probability that LGRB host galaxies and the BCG sample originate from the same parent population is 1\% (or 2\% if we exclude the potentially aberrant data from the AGN host of GRB 031203); comparing the LGRB host galaxies and the SNHGs we find a similar KS test probability of 3\% (or 6\% exclude GRB 031203). By contrast, the LGRB host galaxies and MPGs have a 54\% KS test probability of originating from the same parent sample - this increases to 63\% if we exclude GRB 031203.

To extend this comparison to higher redshifts, we perform a similar comparison between our intermediate-redshift ($0.3 < z < 1.0$) LGRB host sample and the TKRS galaxies (Figure 1, bottom). At these higher redshifts, we are often not able to distinguish between the lower and upper branch metallicities of the R$_{23}$ diagnostic for our LGRB hosts, which poses a challenge to accurate comparisons of metallicities. For this comparison we plot both metallicities in cases where we cannot distinguish between the lower and upper branch. We illustrate the L-Z relation for the TKRS galaxies, as well as L-Z relations assuming both lower-branch and upper-branch R$_{23}$ metallicities for our LGRB host sample where necessary, connected by a dotted line. If the upper-branch metallicities are assumed, we see good agreement between the LGRB host galaxies and the TKRS sample; we find a 48\% KS test probability that these galaxies are drawn from the same parent population. By contrast, for the lower-branch metallicities and L-Z relation we see an offset between the LGRB host galaxies and the TKRS galaxies that is quite similar to our comparison with the BCGs at lower redshift, and we find a much smaller KS test probability of only 0.2\%. If the lower-branch metallicities are correct, this suggests that the low-metallicity bias of LGRB host galaxies may extend out to $z \sim 1$; however, if the upper-branch metallicities are accurate this could imply that LGRB host galaxies at these redshifts are more representative of the general galaxy population than the local sample. Near-infrared observations of these galaxies' optical emission line spectra are required to resolve this R$_{23}$ diagnostic degeneracy and reach robust conclusions regarding the nature of the $z > 0.3$ sample metallicity.

\subsubsection{Metallicity vs. Young Stellar Population Age}
In Figure 2 we compare metallicity to the age of the young stellar population for the $z < 0.3$ LGRB host galaxies, BCGs, and MPGs. Here we can immediately see that the MPGs and BCGs occupy very distinct regions of the diagram. However, we find that the LGRB host galaxies show moderate agreement with both samples. The average young stellar population age of the LGRB hosts is 5.2 $\pm$ 0.2 Myr, compared to 6.2 $\pm$ 0.5 Myr for the BCGs and 4.8 $\pm$ 0.1 Myr for the MPGs. Applying the KS test to the ages for these samples, we find a 16\% probability that the LGRB hosts and BCGs are drawn from the same parent population, and an 11\% probability when comparing the LGRB hosts and MPGs. These percentages are higher if we exclude GRB 031203 from the sample, with a 36\% probability comparing the LGRBs and BCGs, and a 24\% probability comparing the LGRBs and MPGs (although, interestingly, the KS test probability when comparing the BCG and MPG sample ages is only 0.01\%). From these comparisons, we find that age does {\it not} appear to be the primary discriminator between LGRB hosts and the BCG and MPG samples.

\subsubsection{Metallicity vs. Ionization Parameter}
In Figure 3 we compare the LGRB host galaxies, BCGs, MPGs, and SNHGs on a metallicity vs. ionization parameter plot. In this case, the BCG and SNHG samples both appear to have higher metallicities and lower ionization parameters that the MPGs and LGRB hosts, occupying opposing ends of the parameter space. A comparison of the MPG and BCG ionization parameters yields an extremely low KS test probability of 10$^{-3}$\%, a measurement reflected in the strong division seen in Figure 3 between the MPG+LGRB host samples (which share a KS test probability of 63\%) and the BCG+SNHG samples (which share a KS test probability of 42\%). We also calculate low KS test probabilities when comparing the LGRB host galaxy ionization parameters to the BCG and SNHG samples, finding 0.5\% and 3\% respectively. In this respect, we find that our nearby LGRB host sample shows very weak agreement with the general BCG galaxy population, in marked contrast with the SNHG sample. The agreement between the LGRB host sample and MPGs is much more robust. We draw the same conclusions from the K-S test statistics if GRB 031203 is excluded from the sample. 

\subsection{Emission Line Ratio Diagnostic Diagrams}
In addition to comparing the derived ISM properties of our LGRB host galaxies and nearby comparison samples, we also compare the observed emission line flux ratios of our full range of comparison samples to a series of optical emission line ratio diagnostic diagrams. This allows us to directly compare the observed emission line fluxes of these galaxies.

\subsubsection{Nearby (z $< 0.3$) Galaxies}
We use three emission line diagnostic diagrams to compare our five nearby ($z < 0.3$) LGRB host galaxies to the local SDSS, NFGS, BCG, MPG, and SNHG comparison samples. The LGRB sample has a redshift range of $0.009 < z < 0.251$, with a median redshift of $z = 0.11$. The comparison samples range from $0.003 < z < 0.137$, with median redshifts ranging from $0.01 < z < 0.073$. It is important to note that the LGRB host sample has a higher redshift range and median redshift; however, it is appropriate to compare these hosts to comparison samples of galaxies in the local universe rather than an intermediate-redshift sample such as the TKRS galaxies.

{\it [NII]/H$\alpha$ vs. [OIII]/H$\beta$}: [NII]$\lambda$6584/H$\alpha$ correlates strongly with metallicity as well as ionization parameter (Kewley et al.\ 2001, Kewley \& Dopita 2002), while [OIII]$\lambda$5007/H$\beta$ is primarily a measure of ionization parameter with a degenerate dependence on metallicity (Baldwin et al.\ 1981, Kewley et al.\ 2004). The close proximity of the emission lines used in each ratio also renders this diagnostic relatively insensitive to extinction corrections. In Figure 4 we use this diagnostic diagram to compare our nearby LGRB host galaxies to the local comparison samples. We also apply the KS test to the emission line ratios, comparing each of our comparison samples to the $z < 0.3$ LGRB host galaxy sample; results of these calculations are given in Table \ref{tab:ks}. We find that the nearby LGRB host galaxies are {\it not} representative of the general galaxy population, with a very low probability ($<$3\%) that they are from the same parent population as the SDSS, NFGS, BCG, and SNHG host galaxy samples. Instead, the LGRB host galaxies' [OIII]/H$\beta$ ratios are more statistically similar to the MPG sample (97\%), although their metallicity-sensitive [NII]/H$\alpha$ ratios show poor a statistical agreement with the MPG sample as well (5\%); see Table \ref{tab:ks}. This lack of agreement with the general galaxy population is in sharp contrast to the SNHG sample; while the environments hosting Type Ic supernovae appear to be evenly distributed throughout the range of emission line ratios found in the SDSS, NFGS, and BCG comparison samples, the LGRB host galaxies show strong agreement with the MPGs but are not statistically similar to these general populations. This once again affirms our observation in Section 4.1 that the LGRB host galaxies are not representative of the general galaxy population.

{\it [NII]/[OII] vs. [OIII]/[OII]}: The metallicity-sensitive [NII]$\lambda$6584/[OII]$\lambda$3727 ratio and the ionization-parameter-sensitive [OIII]$\lambda$5007/[OII]$\lambda$3727 ratio are a more efficient means of isolating metallicity and ionization parameter than the [NII]/H$\alpha$ vs. [OIII]/H$\beta$ diagnostic. [NII]/[OII] has a minimal dependence on ionization parameter thanks to the similar ionization thresholds of [NII] and [OII] (van Zee et al.\ 1998, Dopita et al.\ 2000). [OIII]/[OII], unlike [OIII]/H$\beta$, does not show a degenerate dependence on metallicity, though a metallicity dependence is still evident (Dopita et al.\ 2000, Kewley \& Dopita 2002). Together, these ratios produce a diagnostic grid with very little degeneracy that clearly separates metallicity and ionization parameter. We show a comparison of these ratios for the nearby LGRB host galaxies and local comparison samples in Figure 5. The nearby LGRB host galaxies once again occupy a distinct parameter space from the general galaxy population (with KS test probabilities of $\le$13\%), and for both the [NII]/[OII] and [OIII]/[OII] diagnostic ratios we see a high statistical correspondence between the LGRB host galaxies and the MPGs (71\% and 44\% respectively; see Table \ref{tab:ks}). Here we can also see that the [OIII]/[OII] ratio is extremely high for two of our LGRB host galaxies (GRB 020903 and GRB 031203); this corresponds to the unusually high ionization parameters of these two galaxies (log $q$ = 8.15 and 8.37, respectively), and in the case of GRB 031203 is likely due in part to a contribution from AGN activity.

{\it [SII]/H$\alpha$ vs. [OIII]/H$\beta$}: Finally, the [SII]$\lambda\lambda$6717,6731/H$\alpha$ line ratio is a useful means of tracing the hardness of the photoionizing spectrum present in a galaxy (Dopita et al.\ 2000). This diagnostic is clearly double-valued; however, it still allows us to compare the properties of different galaxy populations, as well as the predictions of the stellar population synthesis models. In Figure 6 we compare the five nearby LGRB host galaxies with [SII]$\lambda\lambda$6717,6731 emission line flues to the local comparison sample. Our host spectra of GRB 020903 and GRB 030329 do not include sufficient [SII]$\lambda\lambda$6717,6731 detections to determine fluxes for these lines; in these cases we determine upper limits for the [SII]/H$\alpha$ ratio. The small size of our LGRB host sample with definitive [SII] detections precludes us from drawing any strong conclusions, but we do see a very similar distribution to the [NII]/H$\alpha$ vs. [OIII]/H$\beta$ comparison, with the LGRB hosts bridging the gap between the MPGs and the general galaxy populations and showing very poor statistical agreement with the [SII]/H$\alpha$ fluxes for all samples ($\le$2\%; see Table \ref{tab:ks}). In these statistics we exclude the [SII]/H$\alpha$ upper limits determined for GRB 020903 and GRB 030329.

In all cases, when the potentially AGN-contaminated host galaxy of GRB 031203 is excluded from these samples we derive comparable percentages from the KS test and draw the same conclusions about the LGRB host galaxy population. The one notable exception concerns the KS test results comparing [OIII]/[OII] for LGRB host galaxies and MPGs - with the high value for the [OIII]/[OII] ratio in the GRB 031203 host removed, the percent likelihood that these two samples come from the same parent population drops from $\sim$44\% to only $\sim$19\%. This discrepancy is noted in Table \ref{tab:ks}.

\subsubsection{Intermediate-Redshift ($0.3 < z < 1$) Galaxies}
To compare our five higher-redshift LGRB host galaxies (median $z = 0.69$) to the intermediate-redshift comparison sample of galaxies from TKRS (median $z = 0.65$), we employ the R$_{23}$ vs. [OIII]/[OII] diagnostic diagram. In Figure 7 the higher-redshift LGRB hosts and the TKRS sample cover similar ranges in the ratio space, unlike the lower-redshift LGRB hosts when compared with the local sample. However, applying the KS test to the R$_{23}$ and [OIII]/[OII] diagnostic ratios gives a 12\% and 16\% probability, respectively, that the $0.3 < z < 1.0$ LGRB host galaxies and the TKRS sample originate from the same parent population. It must be emphasized that this diagnostic is strongly double-valued, preventing us from drawing any definitive conclusions; however, this comparison does suggest that LGRB host galaxies could be more representative of the general galaxy population at higher redshifts.

\section{Comparison with New Stellar Population Synthesis Models}
\label{grids}
We next compare our LGRB host galaxies with grids of stellar population synthesis models produced by Levesque et al.\ (2009) with the Starburst99 evolutionary synthesis code (Leitherer et al.\ 1999, V\`{a}zquez \& Leitherer 2005) and the Mappings III shock and photoionization code (Binette et al.\ 1985, Sutherland \& Dopita 1993). This suite of models successfully reproduces the spectral properties of star-forming galaxies in most cases, including low metallicity. Levesque et al.\ (2009) adopt the stellar atmospheres of Pauldrach et al.\ (2001) and Hillier \& Miller (1998), both of which adopt detailed NLTE treatments of metal opacities that have not previously been included in stellar population synthesis models. These new atmospheres improve the models' agreement with the general galaxy population, particularly at low metallicities, following the prediction of Kewley et al.\ (2001). Although there are some shortcomings in these models that remain to be addressed (such as a persistent insufficient flux in the higher-energy ionizing spectrum, a problem that becomes more evident at later ages), the models allow one to gain an understanding of the importance of metallicity, ionization parameter, the age of the young stellar population, and star formation history on the emission line ratio diagnostic diagrams. For this work, we adopt model grids that assume a continuous star formation history and a young stellar population age of 5.0Myr (in agreement with the average young stellar population age of our LGRB host galaxies). The grid metallicities range from $Z = 0.1Z_{\odot}$ to $Z = 2Z_{\odot}$ and ionization parameters ranging from $q = 1 \times 10^7$ to $q = 4 \times 10^8$. We also assume an electron density of $n_e = 100$.

In Figure 8 we plot the models and all of our LGRB host galaxies with the required flux ratios on the [NII]/H$\alpha$ vs. [OIII]/H$\beta$ diagnostic diagram. We find 3 of our 6 LGRB host galaxies to be in agreement with the models; however, we find that 5 of the 6 LGRBs agree with models plotted at younger ages (0-3Myr), and Levesque et al.\ (2009) note that the agreement of the models degrades with age, a function of decreasing ionizing flux in the higher-energy regime. The single exception to this is the host galaxy of GRB 031203, which is found to have an unusually high [OIII]/H$\beta$ ratio, indicative of a particularly high ionization parameter. This is consistent with our classification of the GRB 031203 host as an AGN, based on criteria from Kewley et al.\ (2006) that are derived based on these emission line diagnostics.

Similarly, in Figure 9 we show a comparison between the LGRB host galaxies and models on the [NII]/[OII] vs. [OIII]/[OII] diagnostic diagram. We find that 4 of our 5 LGRB host galaxies with these line ratios agree with the models, and that this agreement persists across the full range of model ages; the exception is again GRB 031203, due to its high [OIII]/[OII] ratio. GRB 020903 is found to have a notably high [OIII]/[OII] ratio in Figure 5 as well, but is still found to be in agreement with the Levesque et al.\ (2009) models shown here. However, the high [OIII]/[OII] ratio of the GRB 020903 host and the unusual line ratios from the AGN host of GRB 031203 suggest that adopting a grid which extends to higher ionization parameters, and potentially a grid which adopts treatments of alternative excitation mechanisms such as shocks, might be prudent when modeling the environments of LGRB hosts.

The agreement between the models and the LGRB hosts is less satisfactory in the case of the [SII]/H$\alpha$ vs. [OIII]/H$\beta$ diagnostic diagram (Figure 10), where none of LGRB host line ratios agree with the models. We see similarly poor agreement again in the R$_{23}$ vs. [OIII]/[OII] diagnostic diagram (Figure 11), with only 1 out of our 10 LGRB host galaxies showing agreement with the model grid.  This is, however, consistent with the findings of Levesque et al.\ (2009). In a comparison with the nearby galaxy population, this diagnostic highlighted a shortcoming of the models; specifically, that the ionizing spectrum generated by Starburst99 does not produce sufficient ionizing flux to reproduce the [SII] and [OIII] ratios observed in the nearby population. The lack of agreement between the LGRB hosts and the models suggests that a harder ionizing spectrum is necessary if we wish to accurately model the environments of LGRB host galaxies. 

\section{Discussion and Conclusions}
\label{disc}
In this paper we have presented a sample of ten rest-frame optical spectra of LGRB host galaxies, including eight from recent observations at Keck and Magellan and two from the literature. We have used emission line diagnostics to determine metallicities, ionization parameters, young stellar population ages, and star formation rates for these galaxies. We compare the L-Z relation of our nearby ($z < 0.3$) LGRB host galaxies to our samples of BCGs, MPGs, and SNHGs, and find that LGRB host galaxies appear to fall below the L-Z relation for BCGs, as do the MPGs, while the SNHGs and BCGs follow similar relations. A low-metallicity trend among LGRB host galaxies corresponds to the predictions of stellar evolutionary theory when assuming a single-star progenitor model, and indeed suggests that binary evolution channels for LGRB progenitors may not be necessary (Langer \& Norman 2006). Stellar evolutionary models show that radiation-driven mass-loss rates for massive stars will decrease at lower metallicities (Vink et al.\ 2001), in turn leading to the higher rotational velocities that are a critical component of LGRB progenitors (Kudritzki \& Puls 2000, Meynet \& Maeder 2005) and potentially producing a rotationally-driven wind component. While the exact nature of LGRB progenitors still remain a mystery, their evolution at low metallicities appears to be a key component in their eventual generation of a GRB.

Interestingly, the parameter space occupied by our nearby LGRB host galaxy sample is in contrast to the SNHG sample, which is distributed across the full range of emission line ratios observed in the general SDSS galaxy population. Even with these small samples, the differences in these two samples suggest that the ISM environments that give rise to the progenitors of LGRBs and SNHGs are fundamentally distinct. The broad-lined Type Ic supernovae in this comparison sample are likely distinct events rather than LGRBs where the burst goes unobserved due to the effects of beaming, confirming the findings of Modjaz et al.\ (2008).

It is, however, important to note that these results are limited by small number statistics and require more detailed study in order to quantify the ISM environments of LGRB host galaxies as they compare to MPGs and the general galaxy population. Our nearby sample currently consists of only 5 LGRB host galaxies, which limits the accuracy of our KS tests as well as our comparisons with the general galaxy population samples. It therefore remains to be seen whether or not these conclusions will remain accurate as the number of nearby LGRBs with observable host galaxies continues to grow. It should also be noted that, of the five LGRBs in this sample with $z < 0.3$, four of these (GRBs 980425, 020903, 031203, and 060218) are considerably sub-luminous relative to the general population of LGRBs and may represent a phenomenologically distinct sub-sample of LGRBs (e.g., Soderberg 2004, 2006). As a result, extrapolating conclusions drawn from these nearby LGRB host galaxies to the general population is not advisable until additional observations of more nearby LGRBs can be undertaken.

We also find that one of the galaxies in our nearby sample, the host of GRB 031203, shows evidence of AGN activity. Our statistics analyses demonstrate that including this galaxy in the nearby host sample does not, for the most part, change our conclusions about the population as a whole. However, AGN activity has never been previously observed in LGRB host galaxies, and this could have important implications for the host environments that produce LGRB progenitors. Additional detailed and multi-wavelength observations of the GRB 031203 host galaxy are necessary to learn more about this unusual host environment.

We compare our five higher-redshift ($0.3 < z < 1.0$) LGRB host galaxies to the higher-redshift comparison sample of galaxies from TKRS on the R$_{23}$ vs. [OIII]/[OII] diagnostic diagram. In this case we find agreement between the two populations. Although this diagnostic diagram is double-valued and prevents us from drawing any definitive conclusions, this suggests that, at higher redshifts, LGRB host galaxies might become more representative of the general galaxy population. This is not an unexpected result, as it has been proposed that LGRBs could well be biased towards lower metallicities but still be valid tracers of star formation at high redshift, due to the mean metallicity for galaxies becoming lower at higher redshifts ($Z \approx 0.44 Z_{\odot}$ from Zwaan et al.\ 2005 for $z > 2$). We tentatively suggest that agreement between LGRB hosts and general population could improve further at higher redshifts. However, a larger sample of LGRB host galaxy ISM properties at higher redshifts is critical to further investigate this possibility. Our current analyses are limited by a lack of near-IR spectra for these host galaxies, which are required in order to conclusively determine their metallicities. As in the $z < 0.3$ case, these comparisons and statistics could also be affected by potential unknown biases as a result of the small sample size.

If our results do hold for larger samples, it is worth considering in this discussion that the most commonly-proposed progenitors for LGRBs are thought to be massive Wolf-Rayet stars that are rapidly-rotating and have never evolved through the red supergiant stage, a high-mass-loss phase expected to severely decrease these stars' angular momentum (e.g., Hirschi et al.\ 2005, Yoon et al.\ 2006, Langer \& Norman 2006, Woosley \& Heger 2006). It is also theorized that Wolf-Rayet progenitors of LGRBs are of the evolved carbon-rich (WC) or even oxygen-rich (WO) subtype rather than the earlier nitrogen-rich (WN) subtype. However, this expectation is paradoxical when considering the observations of massive stellar populations at low metallicities. The Wolf-Rayet/red supergiant ratio is actually found to {\it decrease} strongly as a function of decreasing metallicity in Local Group galaxies, with Wolf-Rayet stars becoming rare in low-metallicity Local Group galaxies such as the Small Magellanic Cloud and NGC 6822 (Massey 2003). Furthermore, the ratio of WC/WN Wolf-Rayet stars also decreases with metallicity; evolved Wolf-Rayet stars in particular are found to be very rare in low-metallicity environments (Massey 2003). It is possible that this observed effect of metallicity on the number of late-type Wolf-Rayet stars could impose a lower metallicity limit on environments that could host LGRBs. Alternatively, it is also possible that our existing models for LGRB progenitors still need to be refined to accommodate the effects that low-metallicity environments will have on these stars. Current stellar evolutionary tracks are at odds with the observed Wolf-Rayet/red supergiant and WC/WN ratios, illustrating that our understanding of massive stellar evolution at low metallicities is still limited.

Finally, we compare our full sample of LGRB host galaxies to the stellar population synthesis and photoionization models of Levesque et al.\ (2009). We find that the LGRB hosts and models agree in the case of the [NII]/H$\alpha$ vs. [OIII]/H$\beta$ and [NII]/[OII] vs. [OIII]/[OII] diagnostic diagrams, although decreased agreement at later young stellar population ages of the LGRB hosts is apparent. The models do not reproduce the positions of the LGRB hosts on the [SII]/H$\alpha$ vs. [OIII]/H$\beta$ and R$_{23}$ vs. [OIII]/[OII] diagrams; in Levesque et al.\ (2009), we show that the models produce insufficient ionizing flux to accurately model emission line fluxes such as [SII] and [OIII]. From the poor agreement shown here, it appears that a sufficient hard ionizing spectrum is critical when modeling the environments of LGRB host galaxies.

Additional LGRB host spectra will shed further light on this investigation into the nature of these events' host ISM environments, increasing the sample size for comparison with the other galaxy samples and improving the overall robustness of the statistical analyses. Such observations should also include NIR spectra for intermediate-redshift hosts that will allow more reliable determinations of metallicity. In the future, it would also be beneficial to extend analyses of LGRB host ISM environments to higher redshifts. This would allow a rigorous examination of whether the low-metallicity trend in these host galaxies extends to higher redshifts, and what effect this might have on the utility of LGRBs as tracers of star formation at high redshifts. Finally, a comparison between the host galaxies of LGRBs and short-duration GRBs (Berger 2009) could provide important insights into the progenitors of these enigmatic events.

We gratefully acknowledge useful correspondence with Warren Brown, Andy Fruchter, Margaret Geller, John Graham, Dan Kocevski, Aybuke K\"{u}pc\"{u} Yoldas, Claus Leitherer, Philip Massey, George Meynet, Maryam Modjaz, Daniel Schaerer, and Leonie Snijders regarding this work. We would also like to thank the anonymous referee for constructive and thoughtful comments on this manuscript. We are grateful for the hospitality and assistance of the W. M. Keck Observatories in Hawaii, in particular the guidance and assistance of Greg Wirth, as well as the Las Campanas Observatory in Chile. This paper made use of data from the Gamma-Ray Burst Coordinates Network (GCN) circulars. The authors wish to recognize and acknowledge the very significant cultural role and reverence that the summit of Mauna Kea has always had within the indigenous Hawaiian community.  We are most fortunate to have the opportunity to conduct observations from this sacred mountain. E. Levesque's participation was made possible in part by a Ford Foundation Predoctoral Fellowship. L. Kewley and E. Levesque gratefully acknowledge support by NSF EARLY CAREER AWARD AST07-48559.

\begin{figure}
\epsscale{0.8}
\plotone{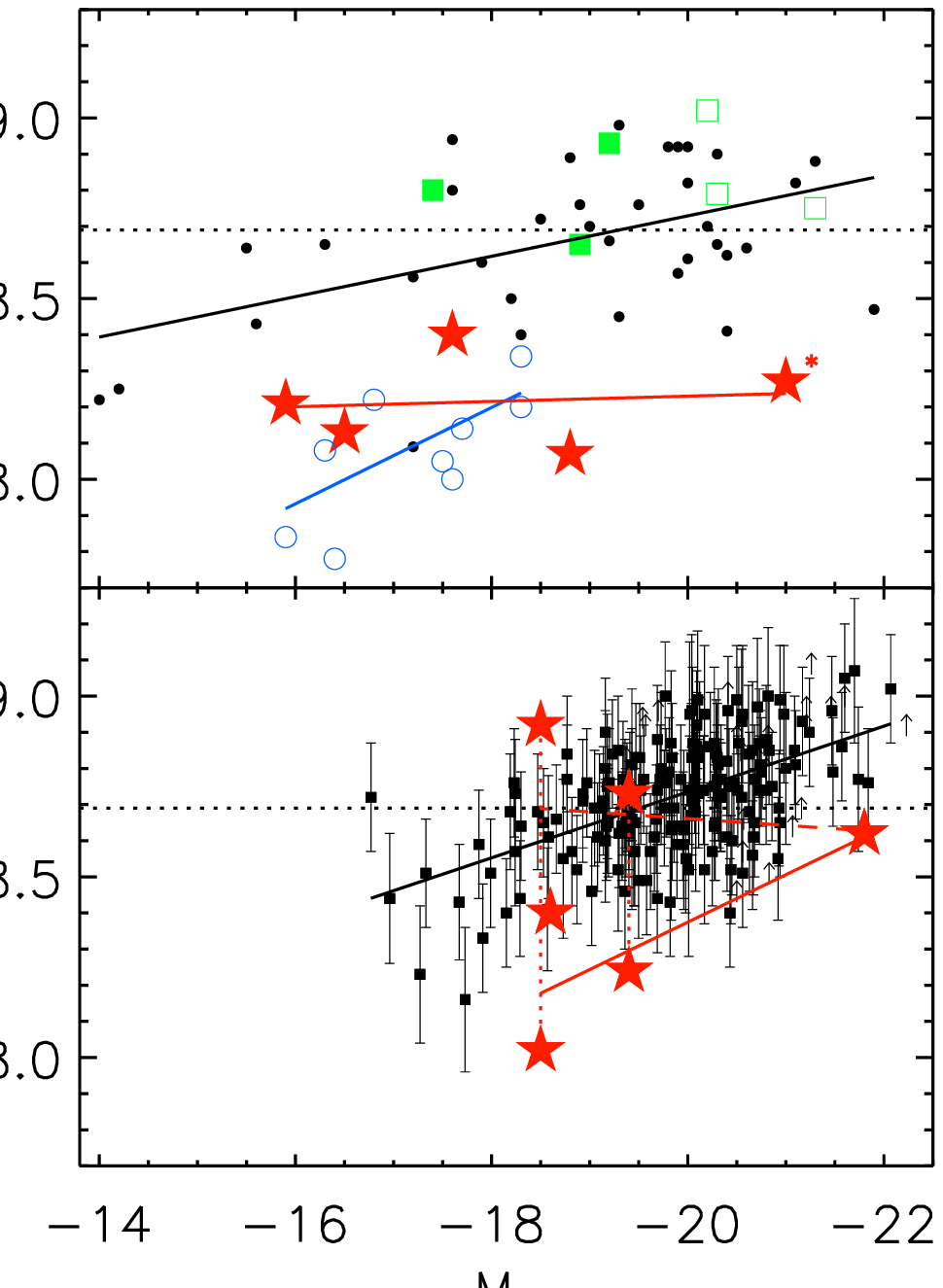}
\caption{Top: Comparison of the luminosity vs. metallicity relations for the BCGs (dots), SNHGs (green squares), MPGs (blue circles), and our low-redshift ($z < 0.3$) LGRB host galaxies (red stars). Solid green squares represent SNHG global galaxy spectra, while open green squares represent SNHG spectra taken at the site of the supernova. The luminosity-metallicity relations for the BCG, MPG, and LGRB host samples are plotted as black, blue, and red solid lines, respectively. GRB 031203, which could potentially be contaminated by contribution from an AGN, is marked with an asterisk. Bottom: Comparison of the luminosity vs. metallicity relation for the TKRS galaxies (black squares) and our higher-redshift ($0.3 < z < 1.0$) LGRB hosts (red stars), based on metallicites for the TKRS sample from Kobulnicky \& Kewley (2004); lower limits on the TKRS metallicities are illustrated by the arrows. For galaxies where an upper or lower branch on the R$_{23}$ diagnostic could not be determined, both metallicities are plotted and connected by a dotted line to illustrate their shared origin from a single galaxy spectrum. The L-Z relation for the TKRS sample is plotted as a solid black line. For the LGRB host sample, we plot the L-Z relation based on both the lower-branch (solid red line) and upper-branch (dashed red line) metallicities. In both cases the solar metallicity is marked by the black dotted line, taken to be log(O/H) + 12 = 8.69 from Asplund et al.\ (2005).}
\end{figure}

\begin{figure}
\epsscale{0.9}
\plotone{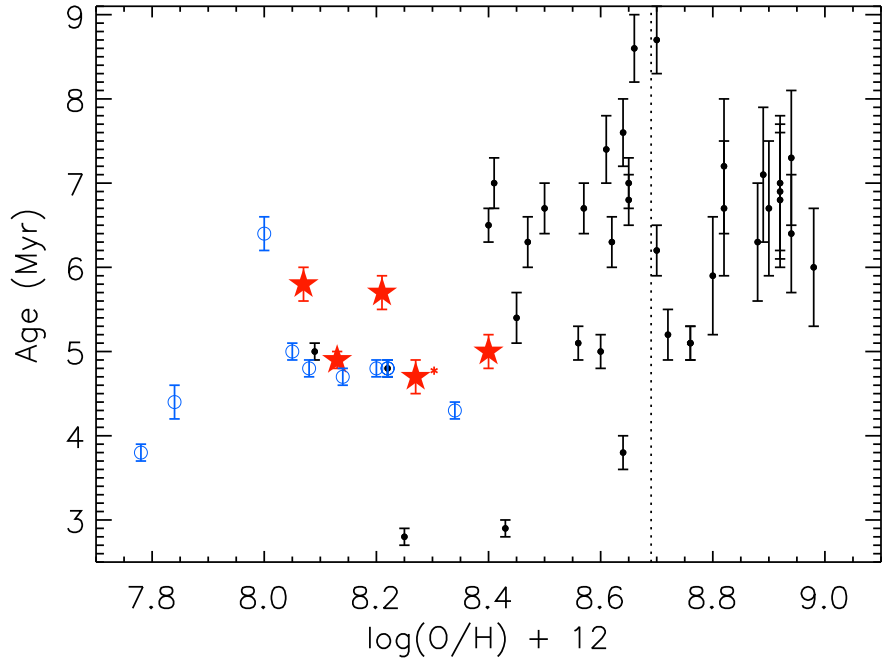}
\caption{Metallicity plotted against young stellar population age for the BCGs (dots), MPGs (blue circles), and our low-redshift ($z < 0.3$) LGRB hosts (red stars). The Asplund et al.\ (2005) solar metallicity log(O/H) + 12 = 8.69 is shown by the black dotted line. GRB 031203 is marked with an asterisk.}
\end{figure}

\begin{figure}
\epsscale{0.9}
\plotone{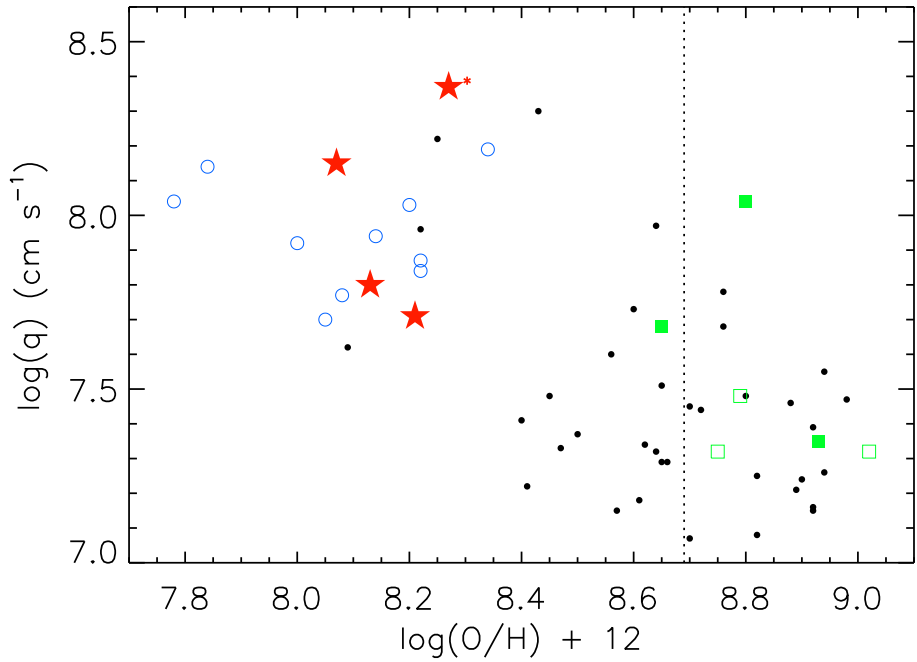}
\caption{Metallicity vs. ionization parameter $q$ for the BCGs (dots), SNHGs (green squares), MPGs (blue circles), and our low-redshift ($z < 0.3$) LGRB host galaxies (red stars). Solid green squares represent SNHG global galaxy spectra, while open green squares represent SNHG spectra taken at the site of the supernova. The Asplund et al.\ (2005) solar metallicity log(O/H) + 12 = 8.69 is shown by the black dotted line. GRB 031203 is marked with an asterisk.}
\end{figure}

\begin{figure}
\epsscale{1}
\plotone{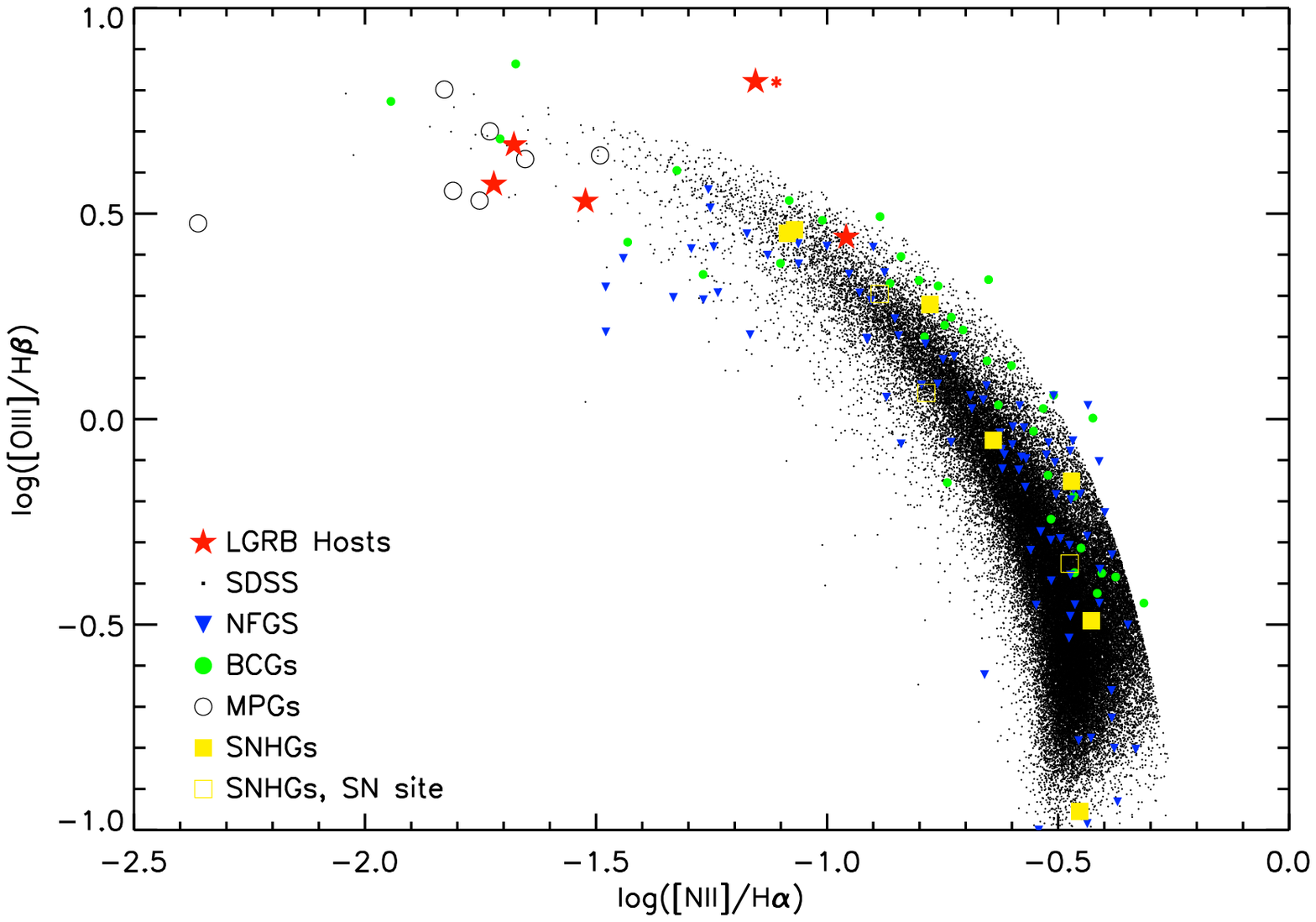}
\caption{Comparison of our low-redshift LGRB host galaxies (red stars) to the SDSS galaxies (points), NFGS galaxies (blue triangles), BCGs (green filled circles), MPGs (large open circles), and SNHGs (yellow squares) on the [NII]/H$\alpha$ vs. [OIII]/H$\beta$ diagnostic diagram. GRB 031203 is marked with an asterisk.}
\end{figure}

\begin{figure}
\epsscale{1}
\plotone{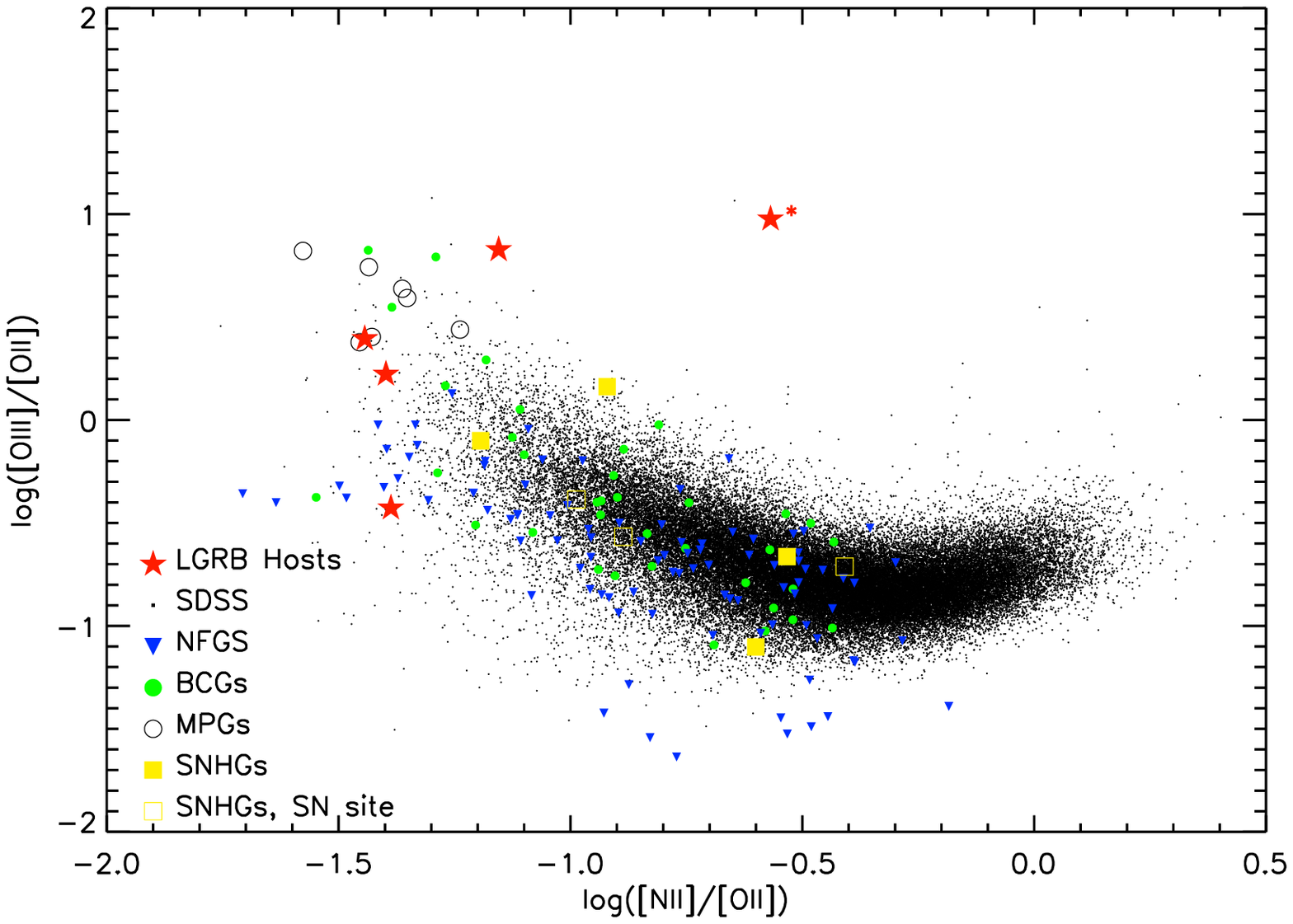}
\caption{Comparison of our low-redshift LGRB host galaxies (red stars) to the SDSS galaxies (points), NFGS galaxies (blue triangles), BCGs (green filled circles), MPGs (large open circles), and SNHGs (yellow squares) on the [NII]/[OII] vs. [OIII]/[OII] diagnostic diagram. GRB 031203 is marked with an asterisk.}
\end{figure}

\begin{figure}
\epsscale{0.98}
\plotone{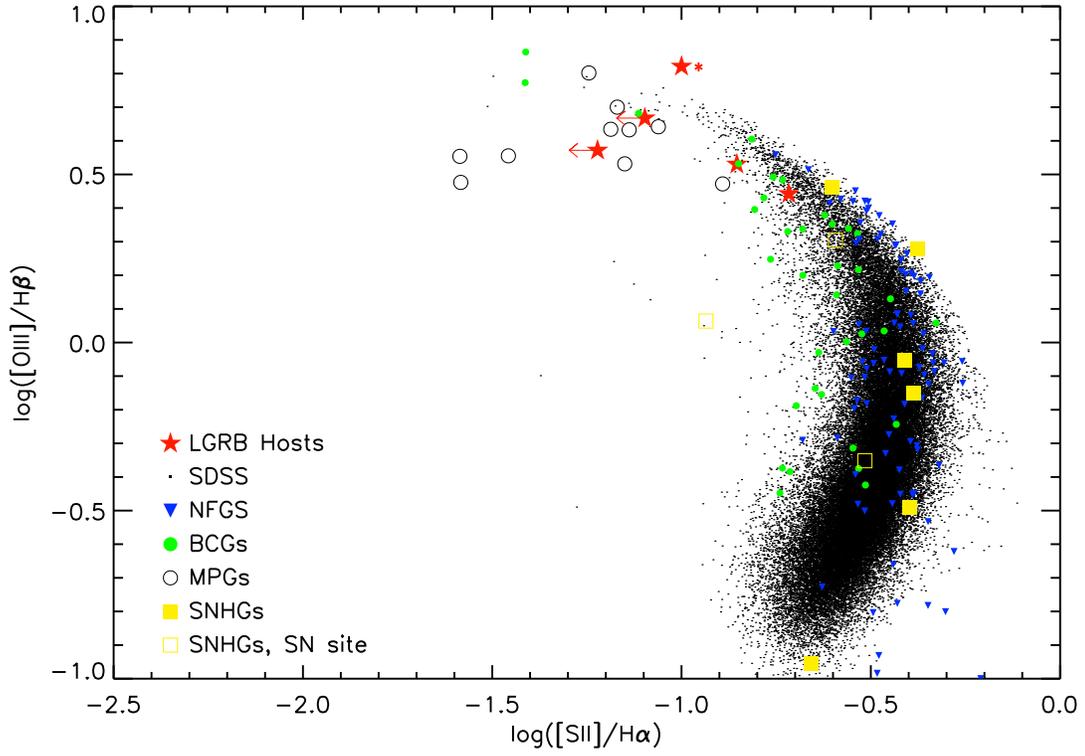}
\caption{Comparison of our low-redshift LGRB host galaxies (red stars) to the SDSS galaxies (points), NFGS galaxies (blue triangles), BCGs (green filled circles), MPGs (large open circles), and SNHGs (yellow squares) on the [SII]/H$\alpha$ vs. [OIII]/H$\beta$ diagnostic diagram. For two of our host galaxies (GRB 020903 and GRB 030329), we are only able to place upper limits on the [SII] line fluxes; the [SII]/H$\alpha$ ratios for these galaxies are therefore denoted as upper limits with arrows. GRB 031203 is marked with an asterisk.}
\end{figure}

\begin{figure}
\epsscale{0.98}
\plotone{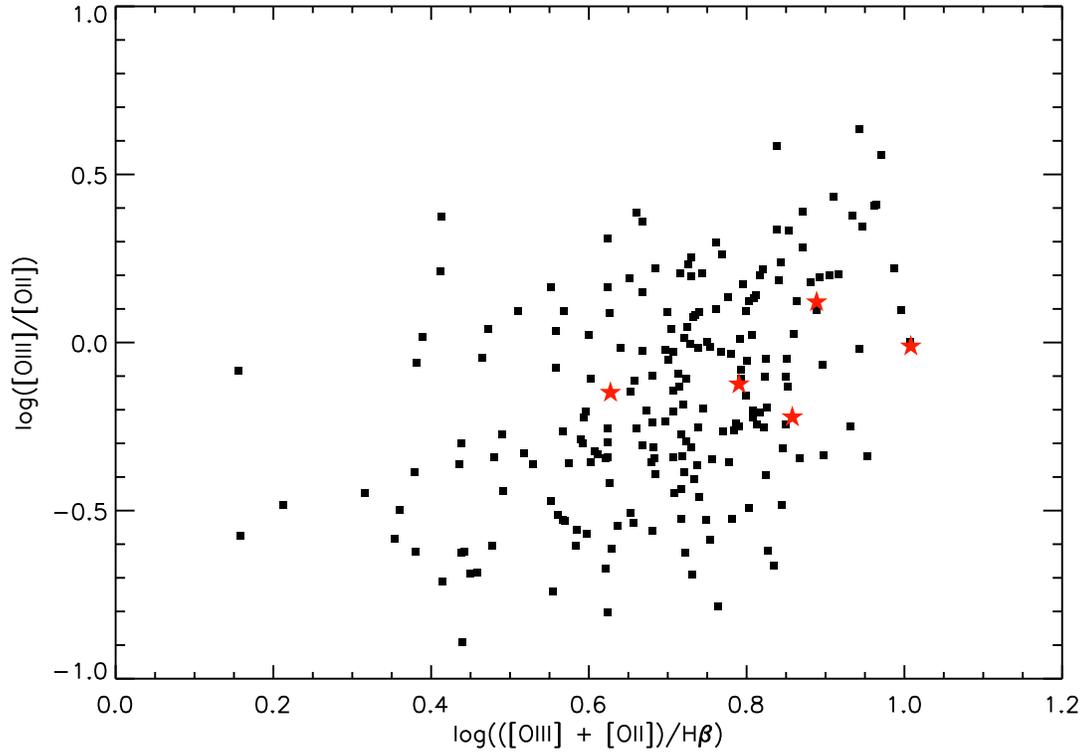}
\caption{Comparison of our high-redshift LGRB host galaxies (red stars) to the TKRS galaxies (black squares) on the R$_{23}$ vs. [OIII]/[OII] diagnostic diagram. On this diagram there appears to be good agreement between the LGRB hosts and the TKRS sample; however, it must be stressed that the R$_{23}$ diagnostic is double-valued, which prevents us from drawing any definitive conclusions regarding the agreement between these two samples.}
\end{figure}

\begin{figure}
\plotone{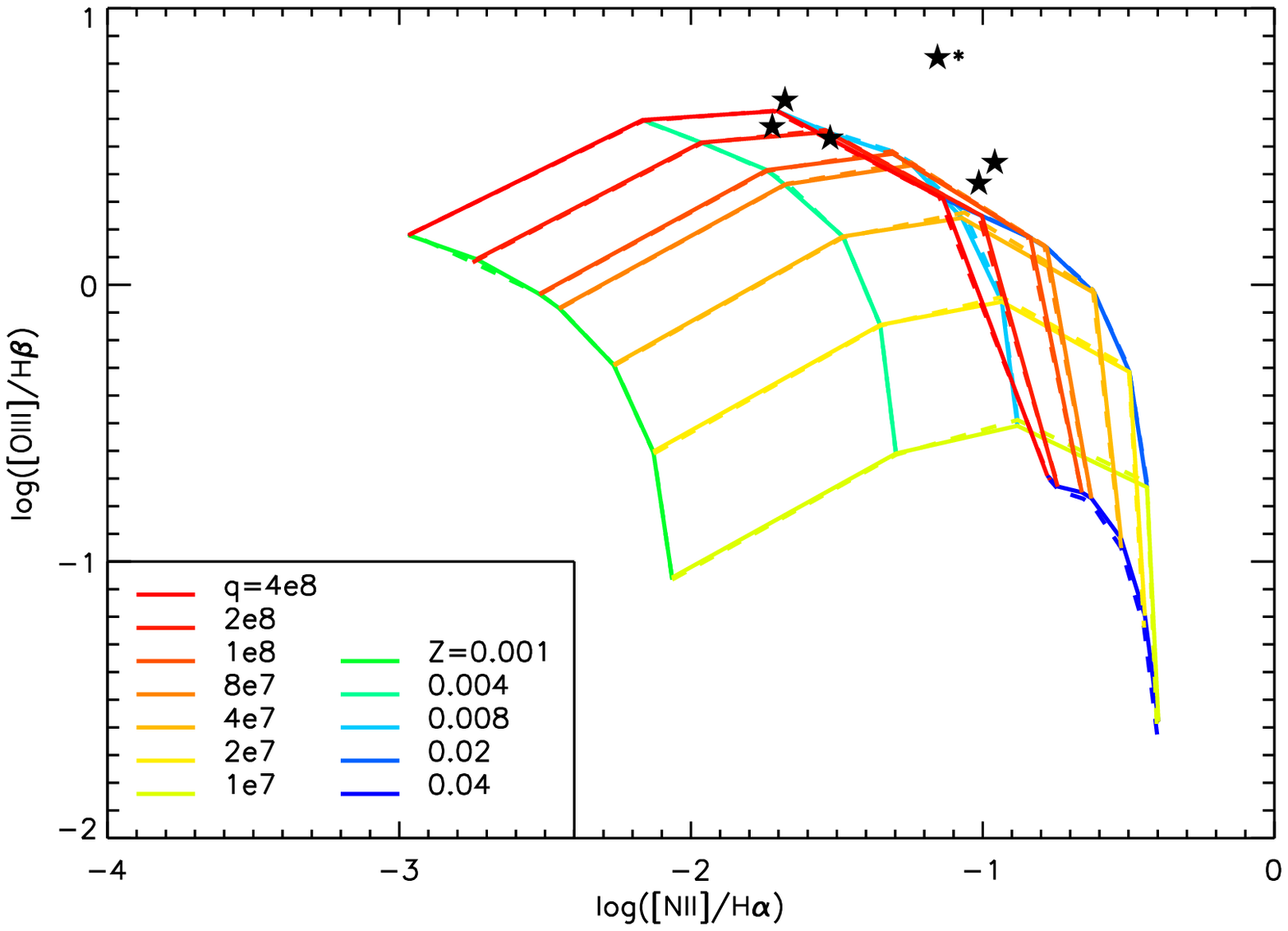}
\caption{Our LGRB host galaxies (stars) compared to the Levesque et al.\ (2009) models on the [NII]/H$\alpha$ vs. [OIII]/H$\beta$ diagnostic diagram. GRB 031203 is marked with an asterisk.}
\end{figure}

\begin{figure}
\plotone{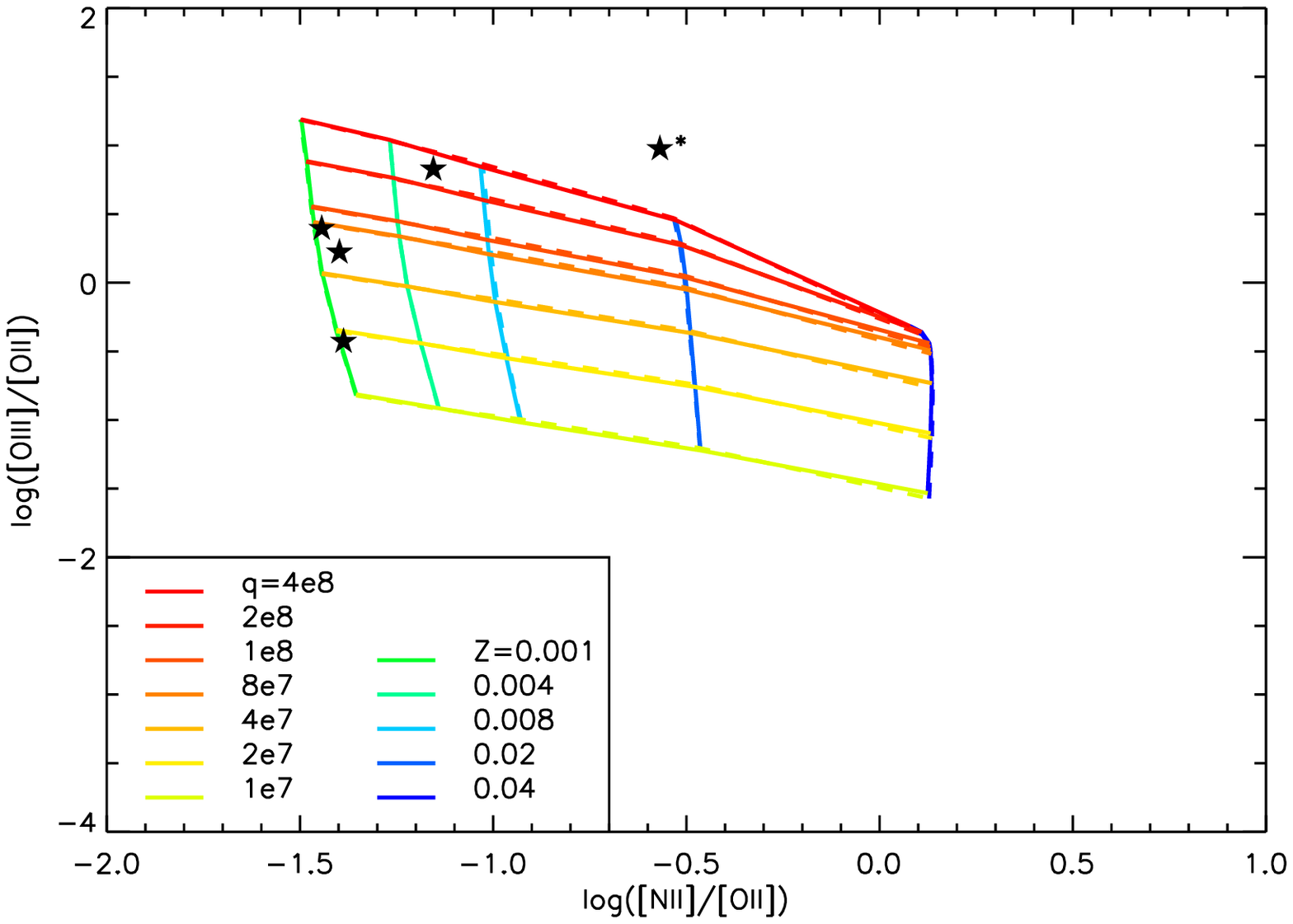}
\caption{Our LGRB host (stars) galaxies compared to the Levesque et al.\ (2009) models on the [NII]/[OII] vs. [OIII]/[OII] diagnostic diagram. GRB 031203 is marked with an asterisk.}
\end{figure}

\begin{figure}
\plotone{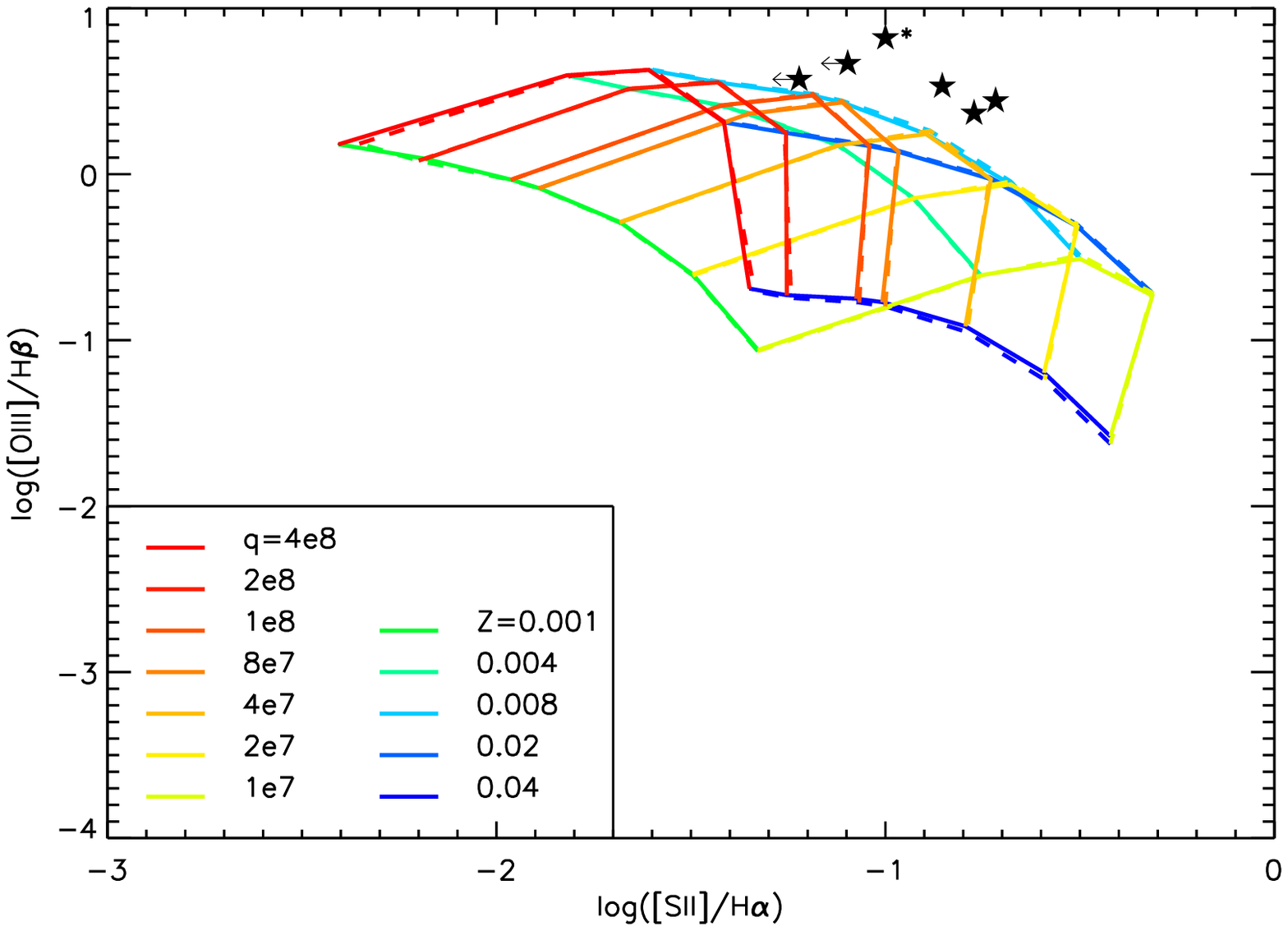}
\caption{Our LGRB host (stars) galaxies compared to the Levesque et al.\ (2009) models on the [SII]/H$\alpha$ vs. [OIII]/H$\beta$ diagnostic diagram. GRB 031203 is marked with an asterisk.}
\end{figure}

\begin{figure}
\plotone{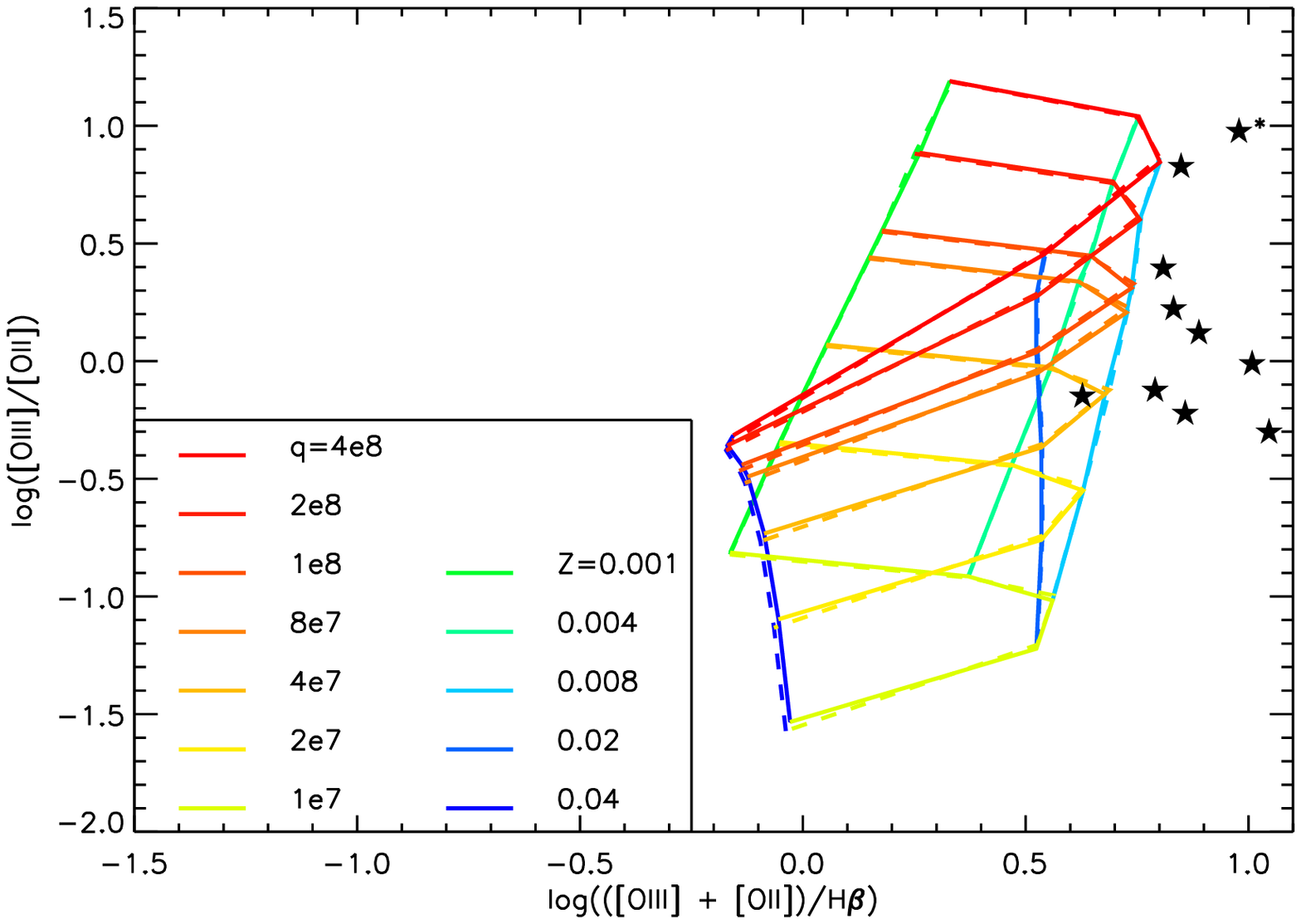}
\caption{Our LGRB host (stars) galaxies compared to the Levesque et al.\ (2009) models on the R$_{23}$ vs. [OIII]/[OII] diagnostic diagram. GRB 031203 is marked with an asterisk.}
\end{figure}

\clearpage
\LongTables
\begin{landscape}
\begin{deluxetable}{l c c c c c c c c}
\tabletypesize{\scriptsize}
\tablewidth{0pc}
\tablenum{1}
\label{tab:params} 
\tablecolumns{9}
\tablecaption{Keck LRIS Observing Set-ups}
\tablehead{
\colhead{Host Galaxy}
&\colhead{$\alpha_{2000}$}
&\colhead{$\delta_{2000}$}
&\colhead{Date (UT)}
&\colhead{Dichroic}
&\colhead{Grism}
&\colhead{Grating}
&\colhead{$\lambda_c$ (\AA)}
&\colhead{Exposure Time}
}
\startdata
GRB 991208 &16 33 53.52 &+46 27 20.88 & 31 May 08 &560 &600/4000 &600/7500 &7400 &5 $\times$ 1800\\
GRB 010921 &22 56 00.00 &+40 55 52.31 & 2 Nov 08 &680 &300/5000 &400/8500 &8100 &4 $\times$ 1800\\
GRB 020903 &22 48 42.23 &$-$20 46 09.12 &7 Oct 03 &560 &400/3400 &400/8500 &7650 &2 $\times$ 900 \\
GRB 031203 &08 02 29.03 &$-$39 51 11.87 &19 Dec 03 &560 &400/3400 &400/8500 &7700 & 2 $\times$ 900 \\
GRB 030329 &10 44 50.00 &+21 31 17.76 &25 Apr 09 &560 &600/4000 &600/7500 &6800 &3 $\times$ 1800 \\
GRB 051022 &23 56 04.10 &+19 35 24.00 & 2 Nov 08 &560 &\nodata &400/8500 &8100 &6 $\times$ 1800\\
GRB 060218 &03 21 39.67 &+16 52 01.56 & 7 Sep 07 &560 &600/4000 &900/5500 &6200 &3 $\times$ 1800\\
\enddata	      	      	
\end{deluxetable}

\begin{deluxetable}{l c c c c c c c c c c c c}
\tabletypesize{\scriptsize}
\tablewidth{0pc}
\tablenum{2}
\label{tab:fluxes}
\tablecolumns{13}
\tablecaption{Diagnostic Emission-Line Fluxes\tablenotemark{a}}
\tablehead{
\colhead{Host Galaxy}
&\colhead{[O II] 3727}
&\colhead{[Ne III] 3869}
&\colhead{H$\delta$ 4101}
&\colhead{H$\gamma$ 4340}
&\colhead{[O III] 4363}
&\colhead{H$\beta$ 4861}
&\colhead{[O III] 4959}
&\colhead{[OIII] 5007}
&\colhead{H$\alpha$ 6563}
&\colhead{[N II] 6584}
&\colhead{[S II] 6717}
&\colhead{[S II] 6730}
}
\startdata
GRB 991208 &0.44 &0.07 &0.06 &0.12 &\nodata &0.34 &0.24 &0.57 &\nodata &\nodata &\nodata &\nodata \\
GRB 010921 &2.74 &0.47 &0.98 &\nodata &\nodata &0.88 &0.60 &2.06 &\nodata &\nodata &\nodata &\nodata \\
GRB 020405 &2.09 &0.31 &0.21 &\nodata &\nodata &0.74 &0.84 &2.77 &\nodata &\nodata &\nodata &\nodata \\
GRB 020903 &3.35 &0.83 &0.52&1.09 &\nodata &4.84 &8.32 &22.5 &11.4 &0.30 &$<$0.32 &$<$0.59 \\
GRB 031203 &0.37 &0.21 &0.15 &0.34 &0.11 &1.57 &3.82 &12.1 &14.3 &1.01 &0.85 &0.67 \\
GRB 030329 &5.95 &1.22 &0.62 &0.88 &0.33 &4.45 &5.44 &16.9 &14.5 &0.28 &$<$0.44 &$<$0.49 \\
GRB 051022 &20.9 &1.67 &1.62 &3.09 &\nodata &8.62 &9.02 &21.4 &\nodata &\nodata &\nodata &\nodata \\
GRB 060218 &8.52 &1.62 &1.16 &2.00 &0.41 &4.24 &5.82 &14.4 &12.3 &0.37 &1.12 &0.66 \\
\enddata	      
\tablenotetext{a} {Raw measured fluxes in units of $10^{-16}$ ergs cm$^2$ s$^{-1}$ \AA$^{-1}$}     	
\end{deluxetable}

\begin{deluxetable}{l c c c c c c c c c c c}
\tabletypesize{\scriptsize}
\tablewidth{0pc}
\tablenum{3}
\label{tab:gals} 
\tablecolumns{11}
\tablecaption{ISM Properties of Our Galaxies}
\tablehead{
\colhead{Galaxy}
&\colhead{$z$}
&\multicolumn{3}{c}{log(O/H) + 12}
&\colhead{log($q$)}
&\colhead{E($B-V$)\tablenotemark{a}}
&\colhead{W$_{H\beta}$\tablenotemark{b}}
&\colhead{Age (Myr)\tablenotemark{c}}
&\colhead{$M_B$ (mag)\tablenotemark{d}}
&\colhead{L(H$\alpha$)}
&\colhead{SFR (M$_{\odot}$/yr)} \\ \cline{3-5}
\multicolumn{2}{c}{}
&\colhead{$T_e$}
&\colhead{R$_{23}$}
&\colhead{PP04}
&\multicolumn{5}{c}{}
&\colhead{$\times 10^{40}$ (ergs s$^{-1}$)}
&\multicolumn{1}{c}{}
}
\startdata
{\bf LGRBs} & & & & & & & & & & & \\
\hline 
GRB 980425 &0.009 &\nodata &$\sim$8.40    &8.28        &\nodata    &0.34 &\nodata         &$\sim$5.0\tablenotemark{e} &-17.6 &7.2 &0.57 \\ %PP04 changed
GRB 990712 &0.434 &\nodata &$\sim$8.40 &\nodata  &\nodata &0.57                                 &\nodata &\nodata &-18.6            &\nodata &10.7\tablenotemark{f} \\ 
GRB 991208 &0.706 &\nodata &8.02/8.92 &\nodata  &7.38/7.80 &0.58                                 &99.8    &4.2 $\pm$ 0.2/3.8 $\pm$ 0.4 &-18.5 &\nodata &3.47/8.35\tablenotemark{f} \\ 
GRB 010921 &0.451 &\nodata &8.24/8.73 &\nodata  &7.44/7.66 &0.00                                 &11.7    &8.0 $\pm$ 0.2/6.8 $\pm$ 0.3 &-19.4 &\nodata &0.70/1.12\tablenotemark{f} \\ 
GRB 020405 &0.691 &\nodata &8.33/8.59 &\nodata  &7.65/7.78 &0.00                                 &25.6   &6.2 $\pm$ 0.2/5.4 $\pm$ 0.3 &\nodata &\nodata &1.61/2.05\tablenotemark{f} \\ 
GRB 020903 &0.251 &\nodata &8.07          &7.98        &8.15          &0.00                                &31.3   &5.8 $\pm$ 0.2 &-18.8 &21.9 &1.7 \\ %PP04 changed
GRB 031203\tablenotemark{g} &0.105 &7.96 &8.27          &8.10        &8.37         &1.17                                 &103.9    &4.7 $\pm$ 0.1 &-21.0 &60.2 &4.8 \\ %PP04 changed
GRB 030329 &0.168 &7.72 &8.13          &8.00        &7.80         &0.13                                 &59.6    &4.9 $\pm$ 0.1 &-16.5 &15.2 &1.2 \\ %PP04 changed
GRB 051022 &0.807 &\nodata &8.62          &8.37        &7.55         &0.50                                 &29.0    &5.2 $\pm$ 0.3 &-21.8 &\nodata &271\tablenotemark{f}\\ %PP04 changed
GRB 060218 &0.034 &7.62 &8.21          &8.07        &7.71         &0.01                                 &33.2   &5.7 $\pm$ 0.2 &-15.9 &0.34&0.03 \\
\hline
{\bf Type Ic hosts} & & & & & & & & & & &  \\
\hline
Central Galaxies & & & & & & & & & & &  \\
\hline 
SN 1997ef &0.0117 &\nodata &\nodata &8.89 &\nodata  &0.52 &\nodata &\nodata &-20.2 &1.2 &0.09 \\ %PP04 fixed
SN 2003jd &0.0188 &\nodata &\nodata &8.54 &\nodata  &0.36 &\nodata &\nodata &-20.3 &35.0 &2.8\\ %PP04 fixed
SN 2005kr &0.1345 &\nodata &8.78 &8.24 &7.98 &0.19 &\nodata &\nodata &-17.4 &2.1 &0.17 \\ %PP04 fixed
SN 2005ks &0.0987 &\nodata &8.92 &8.63 &7.33  &0.31 &\nodata &\nodata &-19.2 &13.8 &1.1 \\ %PP04 fixed
SN 2006nx &0.1370 &\nodata &8.61 &8.24 &7.61  &0.52 &\nodata &\nodata &-18.9 &5.5 &0.44 \\ %PP04 fixed
SN 2006qk &0.0584 &\nodata &8.87 &8.75 &\nodata &0.61 &\nodata &\nodata &-17.9 &7.7 &0.51 \\ %PP04 fixed
SN 2007I &0.0216 &\nodata &\nodata &8.38 &\nodata  &0.37 &\nodata &\nodata &-16.9 &0.30 &0.02 \\ %PP04 fixed
\hline
SN Position & & & & & & & & & & &  \\
\hline
SN 1997ef &0.0117 &\nodata &9.02 &8.69 &7.34 &0.22 &\nodata &\nodata &-20.2 &0.9 &0.07 \\ %PP04 fixed
SN 2003jd &0.0188 &\nodata &8.79 &8.35  &7.48 &0.14 &\nodata &\nodata &-20.3 &0.8 &0.07 \\ %PP04 fixed
SN 2005nb &0.0238 &\nodata &8.75 &8.46  &7.32 &0.34 &\nodata &\nodata &-21.3 &4.3 &0.34 \\ %PP04 fixed
\hline
{\bf MPGs} & & & & & & & & & & &  \\ %PP04 fixed
\hline 
J120955.67+142155.9 &0.078 &7.80 &8.05 &\nodata  &7.70 &0.00 &54.4 &5.0 $\pm$ 0.1 &-17.5 &2.2  &0.18 \\
J123944.58+145612.8 &0.072 &7.73 &8.14 &\nodata  &7.94 &0.03 &91.2 &4.7 $\pm$ 0.1 &-17.7 &4.2  &0.33 \\
J124638.82+350115.1 &0.065 &7.84 &8.22 &8.00 &7.84 &0.07 &70.7 &4.8 $\pm$ 0.1 &-16.8 &3.8 &0.30 \\
J133424.53+592057.0 &0.073 &7.94 &8.22 &8.05  &7.87 &0.00 &72.5 &4.8 $\pm$ 0.1 &-16.8 &5.2 &0.41 \\
J142250.72+514516.5 &0.039 &\nodata &7.84 &\nodata &8.14 &0.00 &95.9 &4.4 $\pm$ 0.2 &-15.9 &1.6  &0.13 \\
J144158.32+291434.2 &0.046 &7.56 &8.08 &8.00  &7.77 &0.00 &76.7 &4.8 $\pm$ 0.1 &-16.3 &0.6 &0.04 \\
J150316.52+111056.9 &0.078 &7.90 &8.20 &7.95 &8.03 &0.00 &80.0 &4.8 $\pm$ 0.1 &-18.3 &15.3 &1.2 \\
J151221.08+054911.2 &0.080 &\nodata &8.00 &7.97  &7.92 &0.00 &61.6 &6.4 $\pm$ 0.2 &-17.6 &4.1 &0.3 \\
J172955.61+534338.8 &0.081 &8.08 &8.34 &7.89 &8.19 &0.12 &158.0 &4.3 $\pm$ 0.1 &-18.3 &20.3 &1.6 \\
J225900.86+141343.5 &0.030 &7.41 &7.78 &7.82 &8.04 &0.00 &134.0 &3.8 $\pm$ 0.1 &-16.4 &1.9 &0.2 \\
\hline
{\bf BCGs} & & & & & & & & & & &  \\ %PP04 fixed
\hline 
iiizw12     & 0.019   &\nodata &8.57 & 8.52   &7.15  &  0.30 &  12.38   &6.7 $\pm$ 0.3   &-19.9 & 20.0      &  1.58  \\
iiizw33     & 0.028   &\nodata &8.64 & 8.37   &7.32  &  0.50 &   8.80   &7.6 $\pm$ 0.4   &-20.6 & 201.4     &  15.91 \\
vzw155      & 0.029   &\nodata &8.70 & 8.72   &7.07  &  0.90 &   5.68   &8.7 $\pm$ 0.4  &-20.2  & 413.2     &  32.64 \\
iiizw43     & 0.014   &\nodata &8.98 & 8.64   &7.47  &  0.68 &  18.53   &6.0 $\pm$ 0.7   &-19.3 & 79.7      &  6.29   \\
iizw40      & 0.003   &\nodata &8.43 & 7.92    &8.30  &  0.07 &  271.1   &2.9 $\pm$ 0.1  &-15.6  & 9.1       &  0.72    \\
mrk5        & 0.003   &\nodata &8.64 & 8.11    &7.97  &  0.05 &  114.1   &3.8 $\pm$ 0.2  &-15.5 & 0.4       &  0.03    \\
viizw156    & 0.012   &\nodata &8.65 & 8.38   &7.51  &  0.42 &  12.21   &6.8 $\pm$ 0.3   &-20.3 & 4.7       &  0.37    \\
haro1       & 0.014   &\nodata &8.82 & 8.54   &7.25  &  0.71 &  12.46   &6.7 $\pm$ 0.8 &-21.1  & 41.3      &  3.27    \\
mrk390      & 0.025   &\nodata &8.42 & 8.42   &7.22  &  0.52 &  11.16   &7.0 $\pm$ 0.3 &-20.4   & 117.9     &  9.31    \\
zw0855      & 0.010   &\nodata &8.76 & 8.35   &7.68  &  0.26 &  56.28   &5.1 $\pm$ 0.2   &-18.9 & 8.9       &  0.70    \\
mrk105      & 0.013   &\nodata &8.94 & 8.56   &7.55  &  0.49 &  15.22   &6.4 $\pm$ 0.7 &-17.6  & 10.9      &  0.86    \\
mrk402      & 0.024   &\nodata &8.45 & 8.29   &7.48  &  0.38 &  25.00   &5.4 $\pm$ 0.3 &-19.3  & 80.2      &  6.34    \\
iizw44      & 0.021   &\nodata &8.89 & 8.77   &7.21  &  0.58 &   9.45   &7.1 $\pm$ 0.8  &-18.8  & 21.2      &  1.68    \\
haro2       & 0.005   &\nodata &8.72 & 8.41   &7.44  &  0.45 &  30.05   &5.2 $\pm$ 0.3  &-18.5 & 13.0      &  1.03    \\
haro3       & 0.003   &\nodata &8.60 & 8.21   &7.73  &  0.15 &  60.67   &5.0 $\pm$ 0.2  &-17.9  & 3.9       &  0.31    \\
haro25      & 0.026   &\nodata &8.77 & 8.33   &7.78  &  0.30 &  42.88   &5.1 $\pm$ 0.2 &-19.5 & 140.3     &  11.08   \\
haro4       & 0.003   &\nodata &8.22 & 7.97   &7.96  &  0.04 &  68.38   &4.8 $\pm$ 0.1 &-14.0  & 0.2       &  0.02    \\
haro29      & 0.001   &\nodata &8.25 & 7.86   &8.22  &  0.06 &  302.0   &2.8 $\pm$ 0.1  &-14.2 & 0.1       &  0.01    \\
mrk215      & 0.020   &\nodata &8.82 & 8.73   &7.08  &  0.80 &   8.60   &7.2 $\pm$ 0.8   &-20.0 & 202.3     &  15.98   \\
haro32      & 0.016   &\nodata &8.62 & 8.61   &7.34  &  0.34 &  15.14   &6.3 $\pm$ 0.3  &-20.4 & 23.1      &  1.82    \\
haro34      & 0.023   &\nodata &8.92 & 8.70   &7.16  &  0.58 &  11.62   &6.8 $\pm$ 0.8  &-20.0  & 60.6      &  4.79    \\
haro35      & 0.026   &\nodata &8.70 & 8.42   &7.45  &  0.26 &  15.76   &6.2 $\pm$ 0.3  &-19.1  & 29.5      &  2.33    \\
haro37      & 0.014   &\nodata &8.80 & 8.48   &7.48  &  0.20 &  20.15   &5.9 $\pm$ 0.7  &-18.6 & 10.3      &  0.81    \\
mrk57       & 0.026   &\nodata &8.61 & 8.55   &7.18  &  0.48 &   9.51   &7.4 $\pm$ 0.4  &-20.0 & 16.0      &  1.26    \\
mrk235      & 0.024   &\nodata &8.92 & 8.61   &7.39  &  0.45 &  11.14   &6.9 $\pm$ 0.8  &-19.9  & 25.1      &  1.98    \\
mrk241      & 0.027   &\nodata &8.94 & 8.69   &7.26  &  0.74 &   8.44   &7.3 $\pm$ 0.8  &-19.2 & 61.0      &  4.82    \\
izw53       & 0.016   &\nodata &8.66 & 8.55    &7.29  &  0.67 &   5.66   &8.6 $\pm$ 0.4  &-19.2 & 7.1       &  0.56    \\
haro39      & 0.009   &\nodata &8.50 & 8.21   &7.37  &  0.37 &  12.52   &6.7 $\pm$ 0.3 &-18.2  & 1.7       &  0.14    \\
iizw70      & 0.004   &\nodata &8.09 & 8.14   &7.62  &  0.00 &  51.82   &5.0 $\pm$ 0.1  &-17.2 & 1.0       &  0.08    \\
izw117      & 0.019   &\nodata &8.90 & 8.64  &7.24  &  0.63 &  12.52   &6.7 $\pm$ 0.8 &-20.3  & 50.3      &  3.97    \\
izw159      & 0.010   &\nodata &8.56 & 8.25   &7.60  &  0.19 &  34.45   &5.1 $\pm$ 0.2  &-17.2 & 6.6       &  0.52    \\
izw191      & 0.019   &\nodata &8.92 & 8.73   &7.15  &  0.59 &   9.92   &7.0 $\pm$ 0.8  &-19.8 & 40.9      &  3.23    \\
ivzw93      & 0.012   &\nodata &8.40 & 8.26   &7.41  &  0.29 &  22.57   &6.5 $\pm$ 0.2  &-18.3 & 7.2       &  0.57    \\
zw2220      & 0.023   &\nodata &8.88 & 8.59   &7.46  &  0.40 &  15.85   &6.3 $\pm$ 0.7 &-21.3  & 62.0      &  4.90    \\
ivzw149     & 0.011   &\nodata &8.47 & 8.41   &7.33  &  0.33 &  15.24   &6.3 $\pm$ 0.3   &-21.9 & 9.3       &  0.73    \\
zw2335      & 0.005   &\nodata &8.65 & 8.50   &7.29  &  0.34 &  11.09   &7.0 $\pm$ 0.3  &-16.3  & 1.2       &  0.09    \\
\enddata	      	
\tablenotetext{a}{Total color excess in the direction of the galaxy, used to correct for the effects of both Galactic and intrinsic extinction.}
\tablenotetext{b}{Rest-frame equivalent widths.}
\tablenotetext{c}{Ages come from the equations derived for the  Schaerer \& Vacca (1998) models relating H$\beta$ equivalent widths and galaxy ages, adopting the R$_{23}$ metallicities.}
\tablenotetext{d} {$M_B$ values come from the literature as follows: Hammer et al.\ 2006 (GRB 980425), Christensen et al.\ 2004 (GRB 990712, GRB 991208, GRB 010921), Soderberg et al.\ 2004 (GRB 020903), Margutti et al.\ 2007 (GRB 031203), Gorosabel et al.\ 2005 (GRB 030329), Castro-Tirado et al.\ 2007 (GRB 051022), Wiersema et al.\ 2007 (GRB 060218), Modjaz et al.\ 2008 (Type Ic hosts), Brown et al.\ 2008 (MPGs), and Kong \& Cheng 2002 (BCGs).}
\tablenotetext{e} {Value from Christensen et al.\ (2008).}
\tablenotetext{f} {SFR determined from the [OII] line flux and the metallicity-dependent relation from Kewley et al.\ (2004).}
\tablenotetext{g} {Since the host of GRB 031203 is not classified as a purely star-forming galaxy, all ISM properties should be taken as approximate, given the potential unknown contribution of AGN activity.}
\end{deluxetable}

\begin{deluxetable}{l c c c c c c c c c c c}
\tabletypesize{\scriptsize}
\tablewidth{0pc}
\tablenum{4}
\label{tab:ages} 
\tablecolumns{12}
\tablecaption{Coefficients for the Schaerer \& Vacca Age-$W_{H\beta}$ Relations}
\tablehead{
\colhead{$Z$ ($Z_{\odot} = 0.02$)}
& \colhead{A}
& \colhead{B}
& \colhead{C ($\times 10^{-4}$)}
& \colhead{D ($\times 10^{-6}$)}
& \colhead{E ($\times 10^{-8}$)}
& \colhead{F ($\times 10^{-11}$)}
& \colhead{G ($\times 10^{-13}$)}
& \colhead{H ($\times 10^{-16}$)}
& \colhead{I ($\times 10^{-19}$)}
& \colhead{J ($\times 10^{-22}$)}
& \colhead{K}
}
\startdata
0.001 & 6.9092 & 0.0103 &-4.2885 &5.3862 & -3.4268 &12.522 &-2.7352 &3.5244 &-2.4669 &0.72156 & 0.0160 \\
0.004 & 7.0507 &-0.0154 &2.5237 &-2.0683 &0.88392 &-2.0336 &0.23925 &-0.11307 & \nodata &\nodata & 0.0127 \\
0.008 & 7.0736 &-0.0293 &9.1706 &-14.306 &12.076 &-59.374 &17.512 &-30.519 &28.965 &-11.540 & 0.0209 \\
0.020 & 6.9548 & -0.0131 &2.4042 &-2.5412 &1.4763 &-4.7918 &0.84861 &-0.73986 &0.22950& \nodata &0.0505 \\
0.040 & 6.9615 & -0.0172 &4.8163 &-9.0032 &9.4962 &-57.944 &20.948 &-44.297 &50.635 &-24.175 & 0.0417 \\
\enddata
\end{deluxetable}

\begin{deluxetable}{l c c c c c}
\tabletypesize{\scriptsize}
\tablewidth{0pc}
\tablenum{5}
\label{tab:ks} 
\tablecolumns{6}
\tablecaption{Kolmogorov-Smirnoff Percentiles for the Nearby ($z < 0.3$) LGRB Host Sample}
\tablehead{
\colhead{Sample}
& \colhead{[NII]/H$\alpha$}
& \colhead{[NII]/[OII]}
& \colhead{[OIII]/H$\beta$}
& \colhead{[OIII]/[OII]}
& \colhead{[SII]/H$\alpha$}
}
\startdata
SDSS &0.002 &0.003 &0.002 &0.002 &0.000 \\
NFGS &1.79 &5.86 &0.011 &0.220 &0.034 \\
BCGs &2.50 &13.1 &0.965 &1.06 &2.01 \\
SNHGs &1.16 &9.08 &1.16 &9.08 &0.829 \\
MPGs &5.18 &70.7 &97.2 &44.3\tablenotemark{a} &0.033 \\
\enddata
\tablenotetext{a}{This value changes significantly, to 19.1, if the potentially AGN-contaminated [OIII]/[OII] ratio for the GRB 031203 host is excluded from the LGRB host sample.}
\end{deluxetable}
\clearpage
\end{landscape}

\appendix
\section{LGRB Host Galaxy Properties}
\subsection{GRB 980425} 
Christensen et al.\ (2008) publish emission line fluxes for the very low-redshift ($z = 0.009$) host galaxy of GRB 980425 that have been corrected for a Galactic extinction of E($B-V$) = 0.059 from Schlegel et al.\ (1998). We find an E($B-V$) = 0.34 based on their published fluxes, which we use to correct for extinction from the host galaxy. While Christensen et al.\ (2008) do not measure a flux for the [OIII]$\lambda$4959 line, we adopt a flux for this line that is 1/3 the flux of the [OIII]$\lambda$5007 line, and thus determine an R$_{23}$ metallicity of log(O/H) + 12 $\sim$ 8.4, at the turnover point of the double-valued R$_{23}$ diagnostic. Adopting the Pettini \& Pagel (2004) $O3N2$ metallicity diagnostic we determine a metallicity of log(O/H) + 12 = 8.28. We adopt the Christensen et al.\ (2008) young stellar population age of 5.0Myr. Finally, we measure SFR = 0.57 M$_{\odot}$/yr for the host galaxy using the Kennicutt (1998) relation and our dereddened H$\alpha$ line flux, slightly higher than Christensen et al.\ (2008)'s value of 0.23 M$_{\odot}$/yr.

\subsection{GRB 990712} 
K\"{u}pc\"{u} Yoldas et al.\ (2006) publish emission fluxes for the $z = 0.434$ host galaxy of GRB 990712 that have been corrected for a foreground extinction of E(B-V) = 0.03 based on Schlegel et al.\ (1998). We find an E($B-V$) = 0.57 based on the ratio of the H$\gamma$ and H$\beta$ line fluxes from their July 2005 spectrum, and use this to correct the published fluxes. Using fluxes uncorrected for the host extinction, K\"{u}pc\"{u} Yoldas et al.\ (2006) calculate a metallicity of log(O/H) + 12 = $\sim$8.3 based on the Kobulnicky \& Kewley (2004) R$_{23}$ lower branch diagnostic. With our corrected fluxes, we determine a metallicity of log(O/H) + 12 $\sim$ 8.4, at the turnover point of the diagnostic. Finally, we can calculate SFR based on both the [OII] flux and the H$\beta$ flux. Assuming a Balmer decrement of 2.87 (Osterbrock 1989) to calculate an H$\alpha$ flux from the H$\beta$ emission line, we find 14.23 M$_{\odot}$/yr from Kennicutt (1998). Using the metallicity-dependent relation for [OII] from Kewley et al.\ (2004), we find 10.7 M$_{\odot}$/yr, in agreement with the extinction-corrected SFR of 10$^{+15}_{-6}$ M$_{\odot}$/yr from K\"{u}pc\"{u} Yoldas et al.\ (2006). With no reported W$_{H\beta}$ we do not determine an age for the young stellar population in this galaxy.

\subsection{GRB 991208} 
The host galaxy of GRB 991208 is at an intermediate redshift of $z = 0.706$ (Figure A1). We calculate an E($B-V$) = 0.58 based on the H$\gamma$/H$\beta$ ratio and correct the fluxes of our observed emission lines accordingly. We apply the R$_{23}$ metallicity diagnostic to this galaxy; however, in the absence of the [NII]$\lambda$6584 and H$\alpha$ lines, which are redshifted into the near-IR, we cannot determine whether the host galaxy lies on the upper or lower branch of the diagnostic. We therefore calculate metallicities for both branches and find log(O/H) + 12 = 8.02 (lower; log $q$ = 7.38) and log(O/H) + 12 = 8.92 (upper; log $q$ = 7.80). We determine a young stellar population age of 4.2 $\pm$ 0.2 Myr for the lower branch metallicity and 3.8 $\pm$ 0.4 Myr for the upper branch metallicity. Finally, using the flux of the [OII] 3727\AA\ line and the Kewley et al.\ (2004) relation we measure 3.47 M$_{\odot}$/yr for the lower branch metallicity and 8.35 M$_{\odot}$/yr for the upper branch metallicity; calculating an H$\alpha$ emission flux based on the H$\beta$ line and a Balmer decrement of 2.87, we find an SFR of 4.23M$_{\odot}$/yr from Kennicutt (1998). We find that the H$\beta$ line in the GRB 991208 host spectrum is asymmetric toward the red, suggesting the possible presence of an inflow in the galaxy.

\subsection{GRB 010921} 
The host galaxy of GRB 010921 is at an intermediate redshift of $z = 0.451$ (Figure A2). The H$\gamma$/H$\beta$ ratio in this galaxy gives us an E($B-V$) = 0, and thus no correction for extinction is applied. We can apply the R$_{23}$ metallicity diagnostic to this galaxy, but cannot determine whether the host galaxy lies on the upper or lower branch of the diagnostic. We calculate metallicities for both branches and find log(O/H) + 12 = 8.24 (lower; log $q$ = 7.44) and log(O/H) + 12 = 8.73 (upper; log $q$ = 7.66). We also determine a young stellar population age of 8.0 $\pm$ 0.2 Myr for the lower branch metallicity, and 6.8 $\pm$ 0.3 Myr for the upper branch metallicity. Using the flux of the [OII] 3727\AA\ line and the Kewley et al.\ (2004) relation we measure 0.70M$_{\odot}$/yr for the lower branch metallicity and 1.12 M$_{\odot}$/yr for the upper branch metallicity; we also determine an H$\alpha$ emission flux based on the H$\beta$ line and a Balmer decrement of 2.87, finding an SFR of 0.52 from Kennicutt (1998).

\subsection{GRB 020405} 
The host galaxy of GRB 020405 is at an intermediate redshift of $z = 0.691$ (Figure A3). The H$\delta$/H$\beta$ ratio in this galaxy gives us an E($B-V$) = 0, and thus no correction for extinction is applied. We are able to apply the R$_{23}$ metallicity diagnostic to this galaxy, but once again lack the redshifted H$\alpha$ and [NII] emission line fluxes necessary to distinguish between the upper and lower branch metallicities. We calculate metallicities for both branches and find log(O/H) + 12 = 8.33 (lower; log $q$ = 7.65) and log(O/H) + 12 = 8.59 (upper; log $q$ = 7.78). We determine a young stellar population age of 6.2 $\pm$ 0.2 Myr for the lower branch metallicity, and 5.4 $\pm$ 0.3 Myr for the upper branch metallicity. Finally, using the flux of the [OII] 3727\AA\ line and the Kewley et al.\ (2004) relation we measure 1.61 M$_{\odot}$/yr for the lower branch metallicity and 2.05 M$_{\odot}$/yr for the upper branch metallicity; calculating an H$\alpha$ emission flux based on the H$\beta$ line and a Balmer decrement of 2.87, we find an SFR of 1.22 from Kennicutt (1998).

\subsection{GRB 020903} 
The host galaxy of GRB 020903 is a low-redshift galaxy in our sample at $z = 0.251$ (Figure A4). The H$\alpha$/H$\beta$ ratio in this galaxy gives us an E($B-V$) = 0, and thus no correction for extinction is applied. We apply the R$_{23}$ metallicity diagnostic and use the [NII]/H$\alpha$ ratio to place this galaxy on the lower branch; this gives us log(O/H) + 12 = 8.07 (log $q$ = 8.15). We also calculate a Pettini \& Pagel (2004) metallicity of 7.98 based on the $O3N2$ diagnostic. We determine a young stellar population age of 5.4 $\pm$ 0.2 Myr. Finally, using the flux of the H$\alpha$ line we measure a SFR of 1.7 M$_{\odot}$/yr based on Kennicutt (1998).

\subsection{GRB 031203}
The host galaxy of GRB 031203 is a low-redshift galaxy in our sample at $z = 0.105$ (Figure A5, top). Based on the Kewley et al.\ (2006) emission line ratio diagnostics for classifying AGN and star-forming galaxies, this host is classified as either a composite or Seyfert galaxy rather than a star-forming galaxy. The presence of AGN activity could potentially contaminate our derived metallicities and other ISM properties. However, applying the diagnostics employed for the star-forming LGRB hosts in our sample, we determine E($B-V$) = 1.17 based on the H$\alpha$/H$\beta$ ratio, which we adopt when correcting for extinction. This high amount of extinction is consistent with the findings of Margutti et al.\ (2007), who determine a Galactic E($B-V$) = 0.72 and an instrinsic E($B-V$) = 0.38; Cobb et al.\ (2004) also show that GRB 031203's location is very close to the Galactic plane, suggesting that we would expect an increased amount of extinction towards the host galaxy. We apply the R$_{23}$ metallicity diagnostic and detection of the auroral [OIII]$\lambda$4363 line (Figure A5, bottom) to place this galaxy on the lower branch, giving a metallicity of log(O/H) + 12 = 8.27 (log $q$ = 8.37). We also calculate a Pettini \& Pagel (2004) metallicity of 8.10 based on the $O3N2$ diagnostic and log(O/H) + 12 = 7.96 based on $T_e$. We determine a young stellar population age of 4.7 $\pm$ 0.1 Myr. Finally, using the flux of the H$\alpha$ line we measure a SFR of 4.8 M$_{\odot}$/yr based on Kennicutt (1998).

\subsection{GRB 030329} 
The host galaxy of GRB 030329 is a low-redshift galaxy at $z = 0.168$ (Figure A6, top). We determine E($B-V$) = 0.13 based on the H$\alpha$/H$\beta$ ratio, which we adopt when correction for extinction. We apply the R$_{23}$ metallicity diangostic and use the [NII]/[OII] and [NII]/H$\alpha$ ratio along with detection of the [OIII] $\lambda$4363 (Figure A6, bottom) line to place this galaxy on the lower branch, giving us log(O/H) + 12 = 8.13 (log $q$ = 7.80). We also calculate a Pettini \& Pagel (2004) metallicity of log(O/H) + 12 = 8.00 based on the $O3N2$ diagnostic and a $T_e$ metallicity of log(O/H) + 12 = 7.72. We determine a young stellar population age of 4.9 $\pm$ 0.1 Myr. Using the flux of the H$\alpha$ line we measure a SFR of 1.2 M$_{\odot}$/yr based on Kennicutt (1998). We find that the neutral hydrogen lines are double-peaked, indicating two components.

\subsection{GRB 051022} 
The host galaxy of GRB 051022 is at an intermediate redshift of $z = 0.807$ (Figure A7). Based on the H$\gamma$/H$\beta$ ratio we calculate an E($B-V$) = 0.50, which we use to correct the emission line fluxes for extinction. We apply the R$_{23}$ metallicity diagnostic and determine [NII]/H$\alpha$ ratio using equivalent widths from the NIRSPEC near-IR host spectrum of Graham et al.\ (2009a,b). Based on [NII]/H$\alpha$ we place this galaxy on the upper branch of the R$_{23}$ diagnostic; this gives us log(O/H) + 12 = 8.62 (log $q$ = 7.54), a value comparable to the log(O/H) + 12 = 8.77 measured by Graham et al. (2009a) based on the R$_{23}$ diagnostic. Our Pettini \& Pagel (2004) metallicity based on the $O3N2$ diagnostic gives a metallicity of log(O/H) + 12 = 8.37. We determine a young stellar population age of 5.2 $\pm$ 0.3 Myr.

Finally, using the extinction-corrected flux of the [OII] line we measure a surprisingly high SFR of 271 M$_{\odot}$/yr based on Kewley et al.\ (2004). We also determine SFR by using the H$\beta$ emission line flux and the Balmer decrement of 2.87 to estimate the H$\alpha$ flux for this host galaxy. With this estimated flux and the Kennicutt (1998) relation, we determine a comparable SFR of 328 M$_{\odot}$/yr. This high star formation rate is in agreement with the findings of Berger et al.\ (2003), who propose that $\sim$20\% of GRB host galaxies have very high star formation rates on the order of 500 M$_{\odot}$/yr based on sub-millimeter and radio observations of GRB host galaxies.

\subsection{GRB 060218} 
The host galaxy of GRB 060218 is a low-redshift galaxy in our sample at $z = 0.034$ (Figure A8). We determine a very low E($B-V$) = 0.01 based on the H$\alpha$/H$\beta$ ratio, which we adopt when correcting for extinction. We apply the R$_{23}$ metallicity diagnostic and use the [NII]/H$\alpha$ ratio along with detection of the [OIII] $\lambda$4363 line to place this galaxy on the lower branch; this gives us log(O/H) + 12 = 8.21 (log $q$ = 7.71). We also calculate a Pettini \& Pagel (2004) metallicity of 8.07 based on the $O3N2$ diagnostic and a metallicity of 8.30 based on $T_e$. We determine a young stellar population age of 5.6 $\pm$ 0.2 Myr. Finally, using the flux of the H$\alpha$ line we measure a SFR of 0.03 M$_{\odot}$/yr based on Kennicutt (1998). We find that the H$\alpha$ line is slightly asymmetric towards the red, indicating the possible presence of an inflow in the host galaxy.

\begin{figure}
%\plotone{/Volumes/APHID/Diss/Observing/Data/GRBs/GRB991208/LRIS/GRB991208_optical.eps}
\plotone{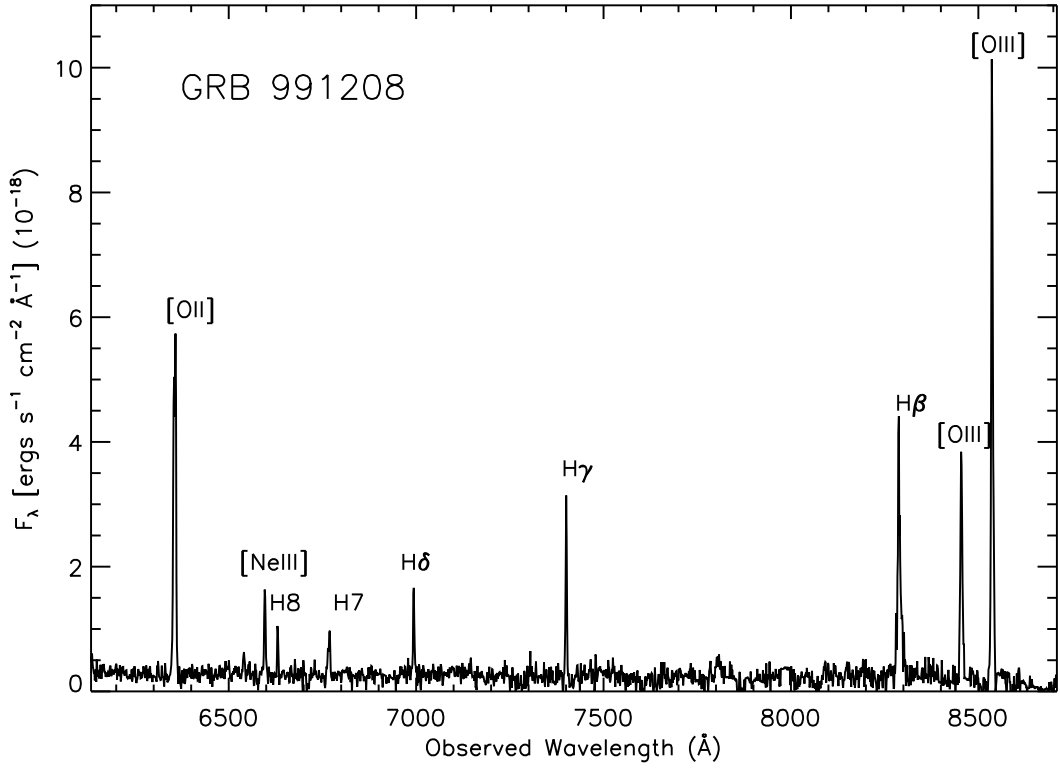}
\caption{Our spectrum of the $z = 0.706$ host galaxy of GRB 991208, observed with LRIS at Keck on 31 May 2008.}
\end{figure}

\begin{figure}
%\plotone{/Volumes/APHID/Diss/Observing/Data/GRBs/GRB010921/LRIS/Nov08/GRB010921_optical.eps}
\plotone{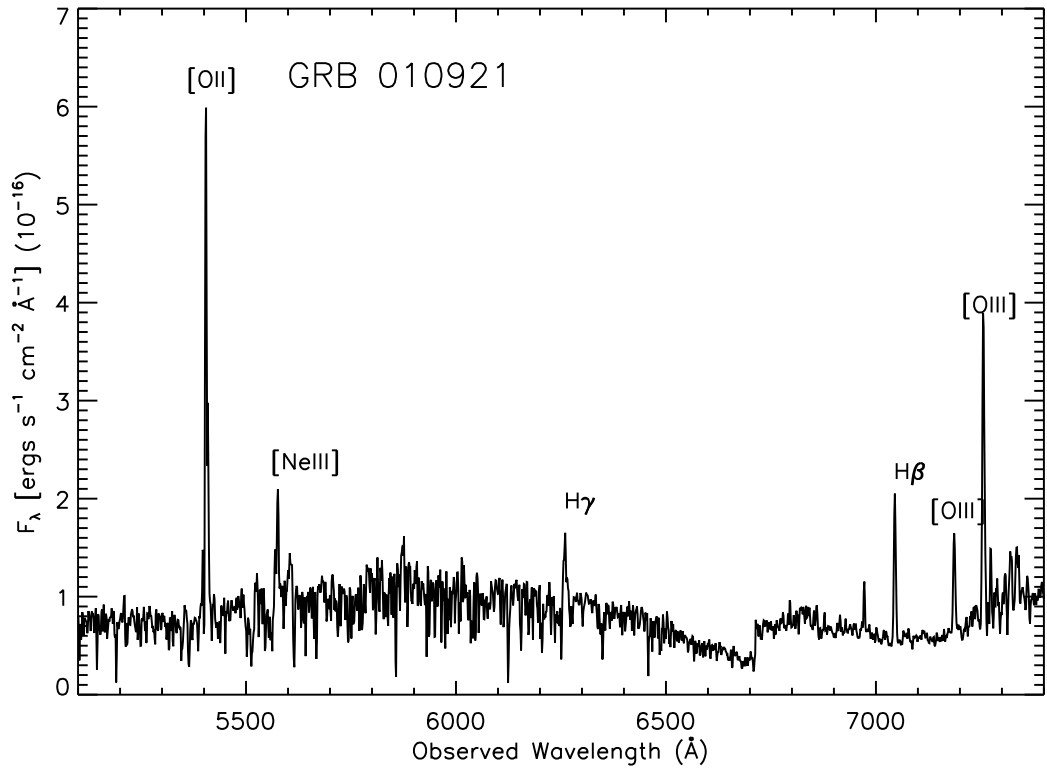}
\caption{Our spectrum of the $z = 0.451$ host galaxy of GRB 010921, observed with LRIS at Keck on 2 November 2008.}
\end{figure}

\begin{figure}
%\plotone{/Volumes/APHID/Diss/Observing/Data/GRBs/GRB020405/GRB020405_optical.eps}
\plotone{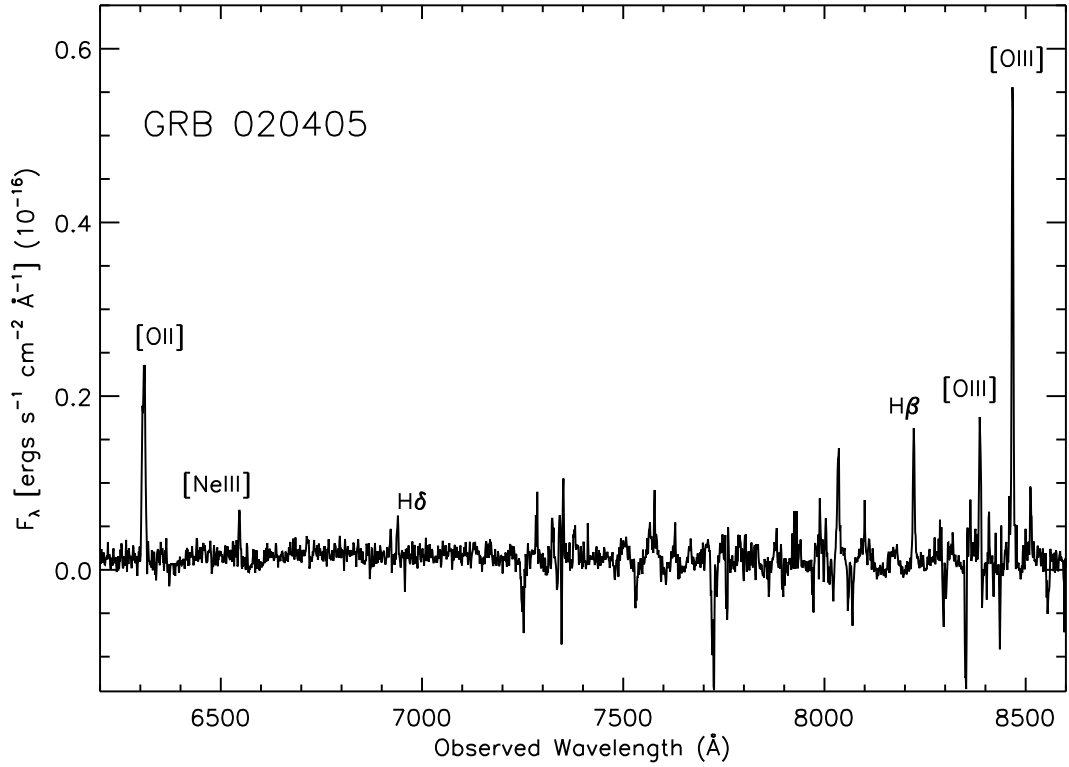}
\caption{Our spectrum of the $z = 0.691$ host galaxy of GRB 020405, observed with LDSS3 at Magellan on 16 January 2008.}
\end{figure}

\begin{figure}
%\plotone{/Volumes/APHID/Diss/Observing/Data/GRBs/GRB020903/Edo/GRB020903_optical.eps}
\plotone{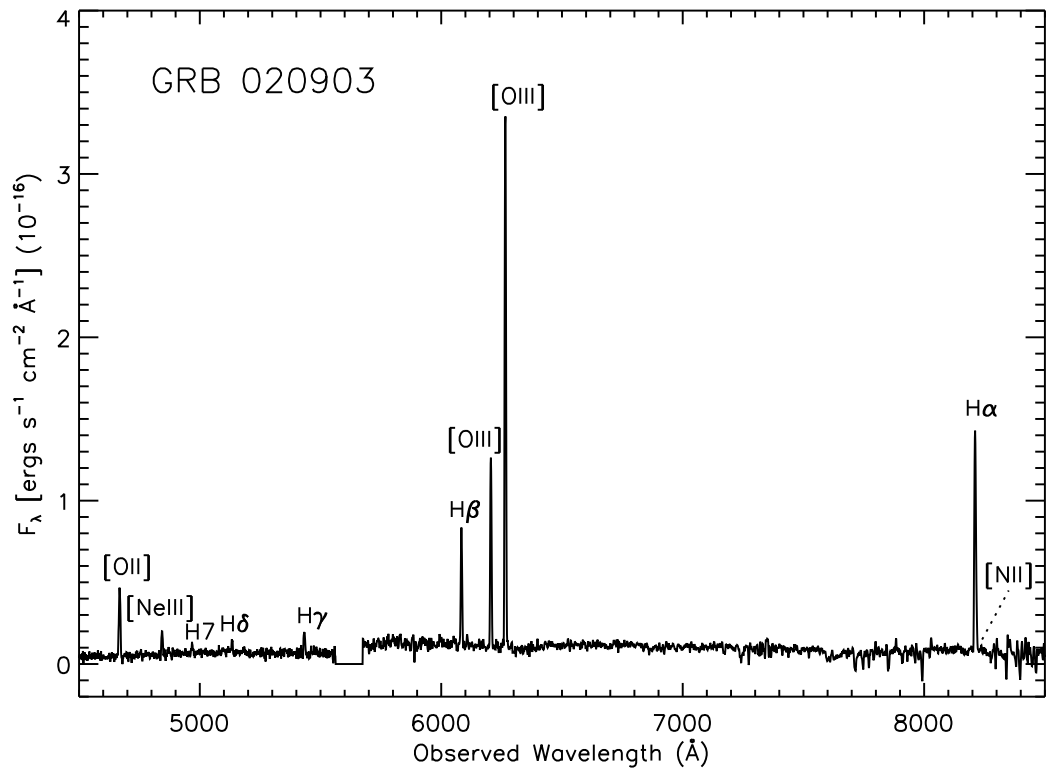}
\caption{Our spectrum of the $z = 0.251$ host galaxy of GRB 020903, observed with LRIS at Keck 7 October 2003}
\end{figure}

\begin{figure}
\epsscale{0.8}
%\plotone{/Volumes/APHID/Diss/Observing/Data/GRBs/GRB031203/Edo/GRB031203_optical.eps}
%\plotone{/Volumes/APHID/Diss/Observing/Data/GRBs/GRB031203/Edo/GRB031203_optical_blue.eps}
\plotone{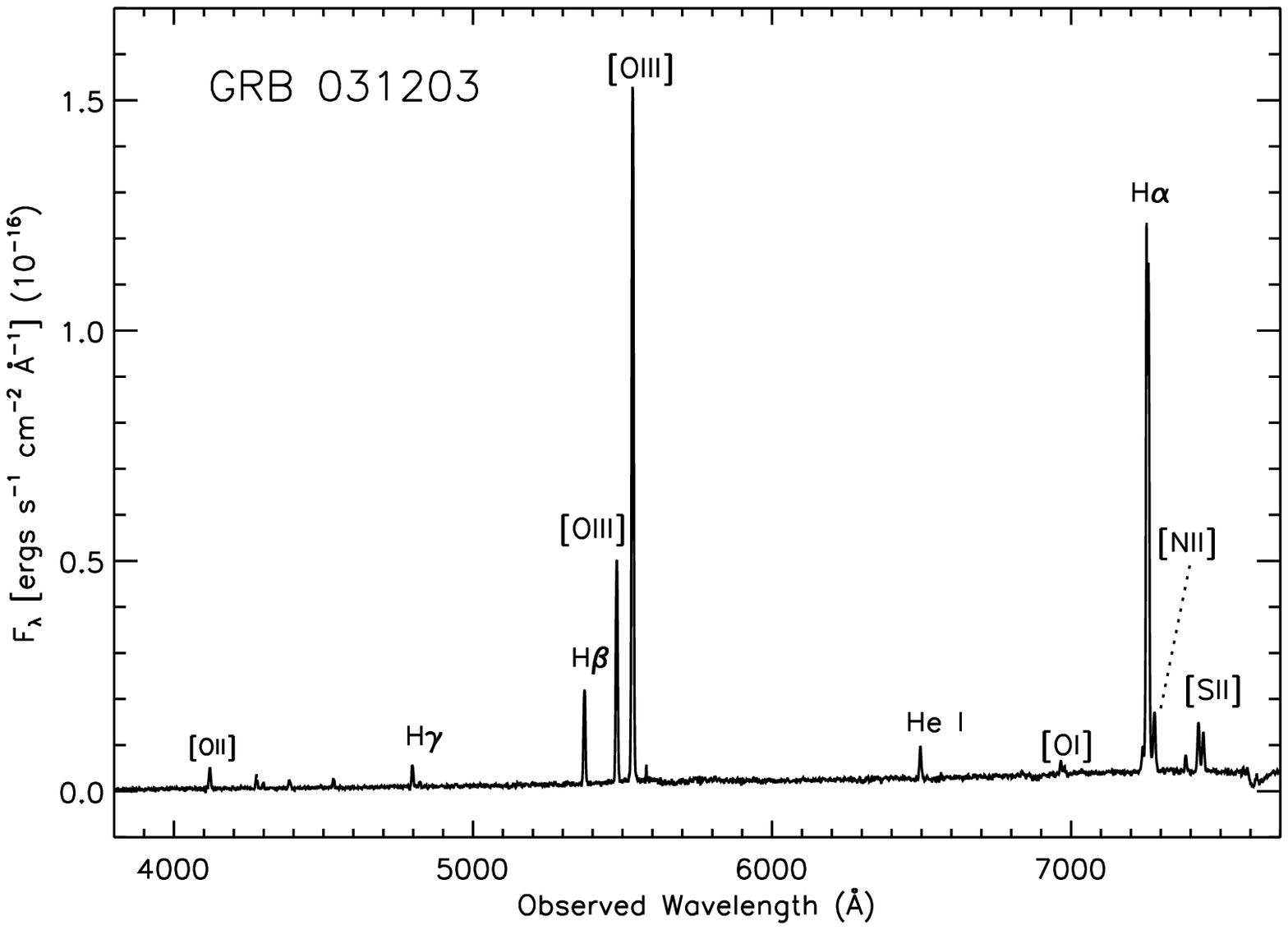}
\plotone{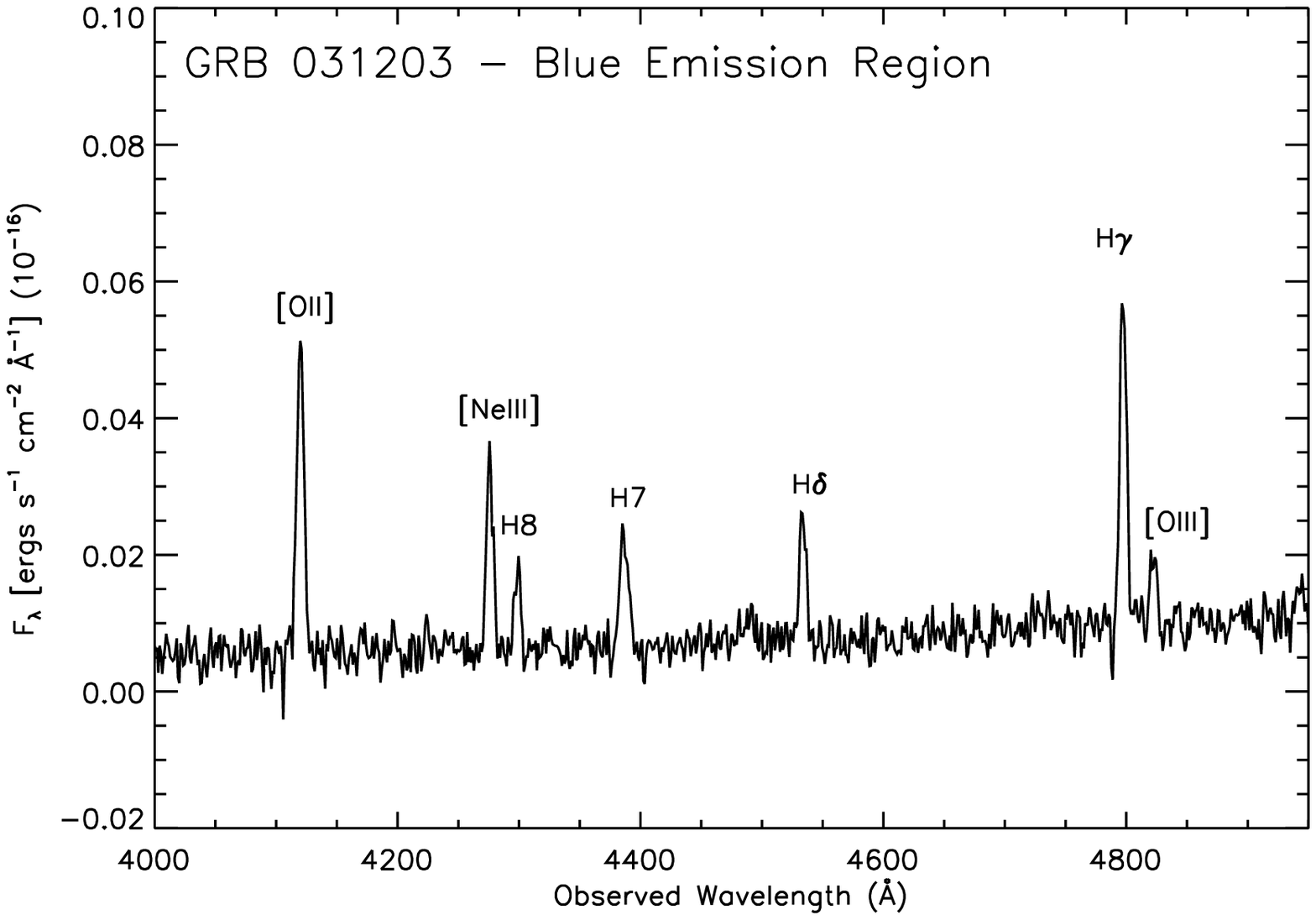}
\caption{Our spectrum of the $z = 0.105$ host galaxy of GRB 031203 observed with LRIS at Keck on 19 December 2003; we plot the entire spectrum(top), with the blue region of the spectrum enhanced to illustrate detection of the [OIII]$\lambda$4363 line (bottom).}
\end{figure}

\begin{figure}
\epsscale{0.8}
%\plotone{/Volumes/APHID/Diss/Observing/Data/GRBs/GRB030329/LRIS/GRB030329_optical.eps}
%\plotone{/Volumes/APHID/Diss/Observing/Data/GRBs/GRB030329/LRIS/GRB030329_optical_zoom.eps}
\plotone{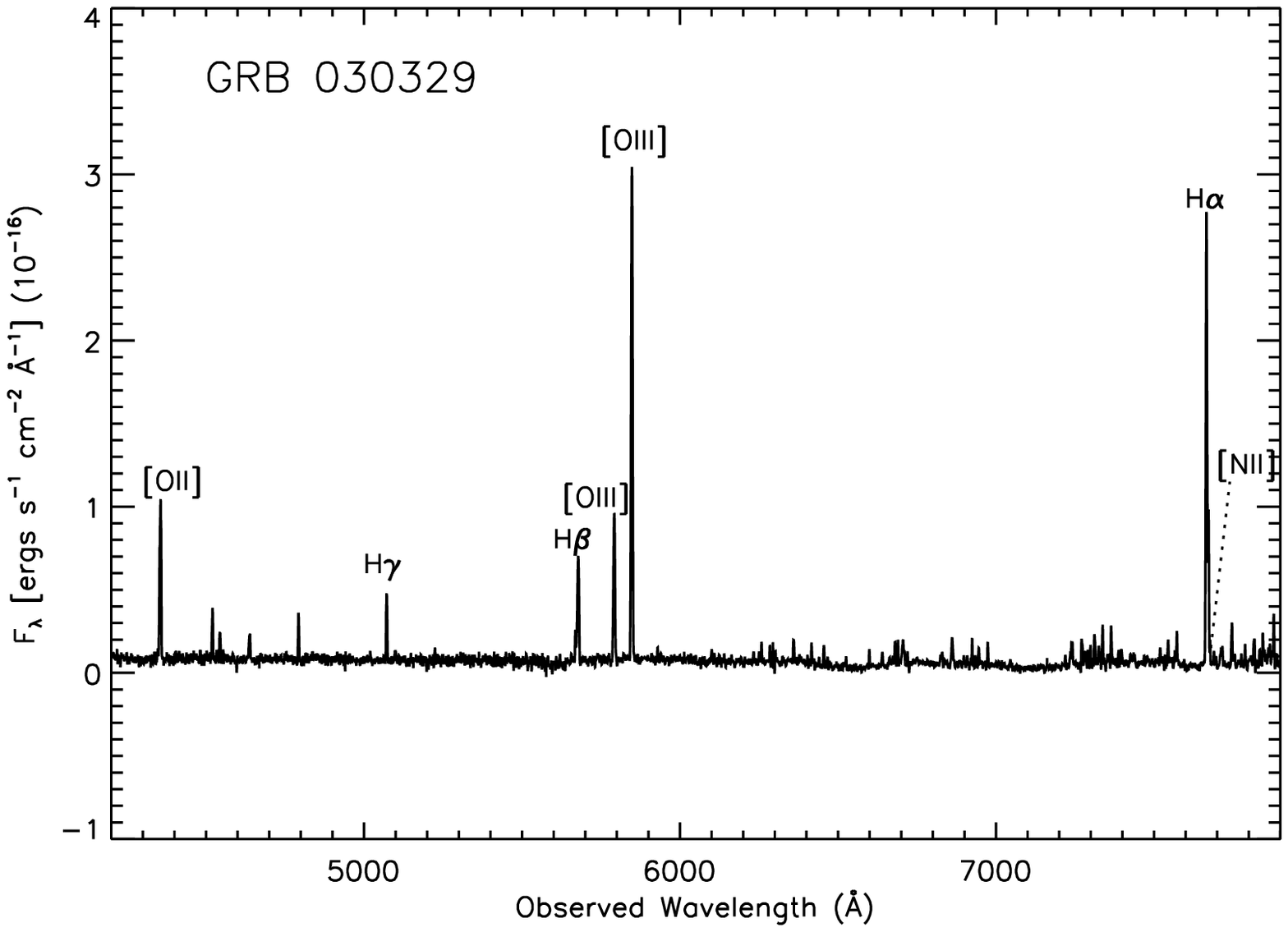}
\plotone{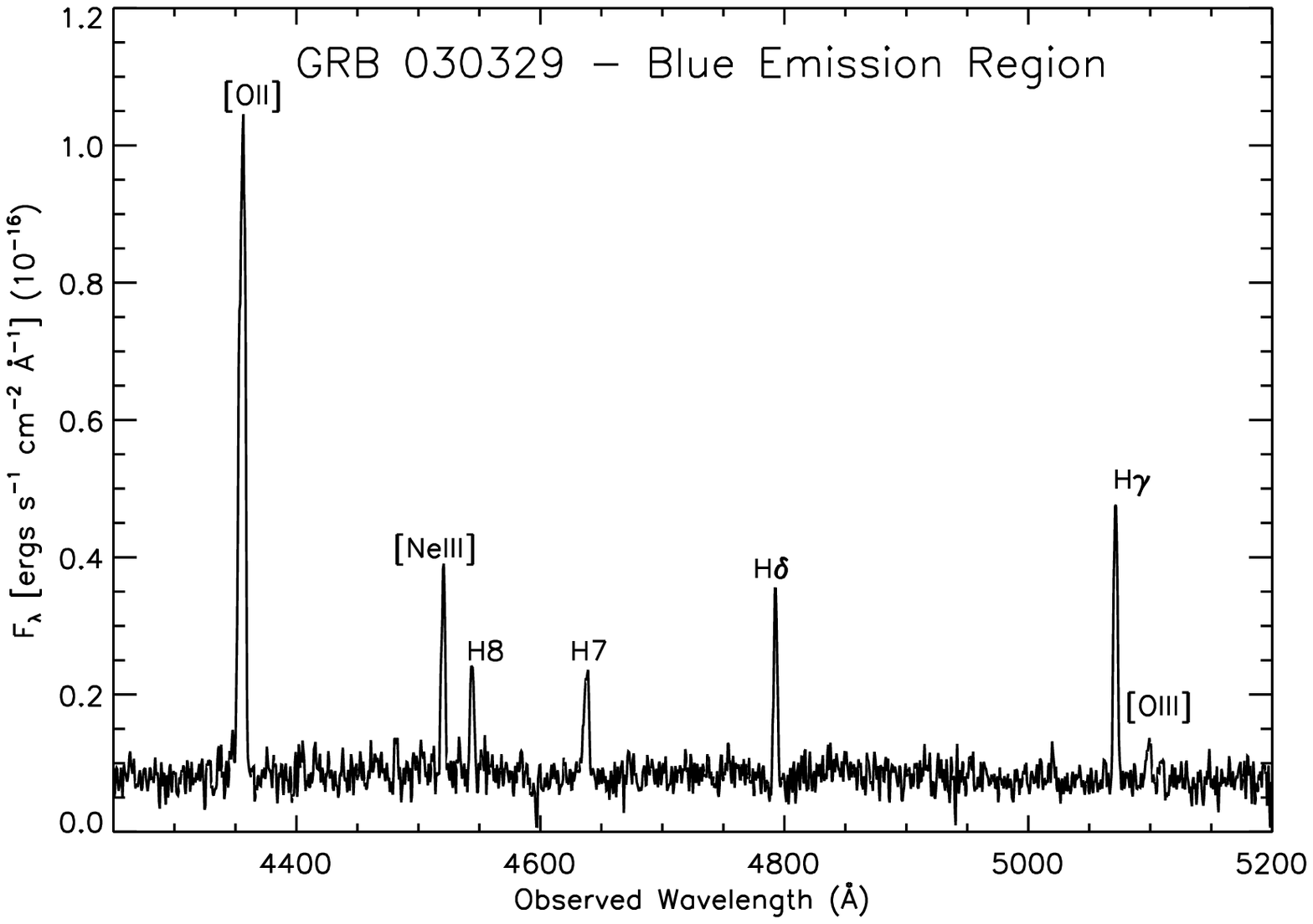}
\caption{Our spectrum of the $z = 0.168$ host galaxy of GRB 030329 observed with LRIS at Keck on 25 April 2009; we plot the entire spectrum (top), with the blue region of the spectrum enhanced to illustrate detection of the [OIII]$\lambda$4363 line (bottom).}
\end{figure}

\begin{figure}
%\plotone{/Volumes/APHID/Diss/Observing/Data/GRBs/GRB051022/LRIS/FINAL_REDUCTON/GRB051022_optical.eps}
%\plotone{/Volumes/APHID/Diss/Observing/Data/GRBs/GRB051022/NIRSPEC/GRB051022_IR.eps}
\plotone{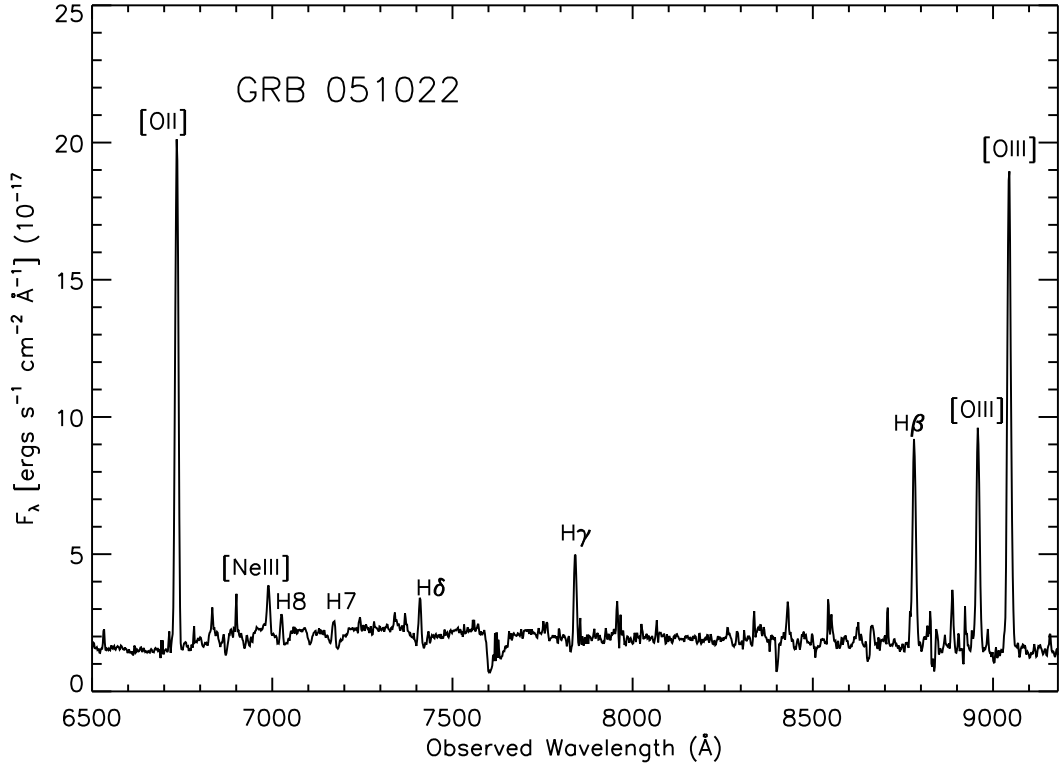}
\caption{Our spectrum of the $z = 0.807$ host galaxy of GRB 051022, observed with LRIS at Keck in the optical on 2 November 2008.}
\end{figure}

\begin{figure}
\epsscale{1}
%\plotone{/Volumes/APHID/Diss/Observing/Data/GRBs/GRB060218/LRIS/GRB060218_optical.eps}
\plotone{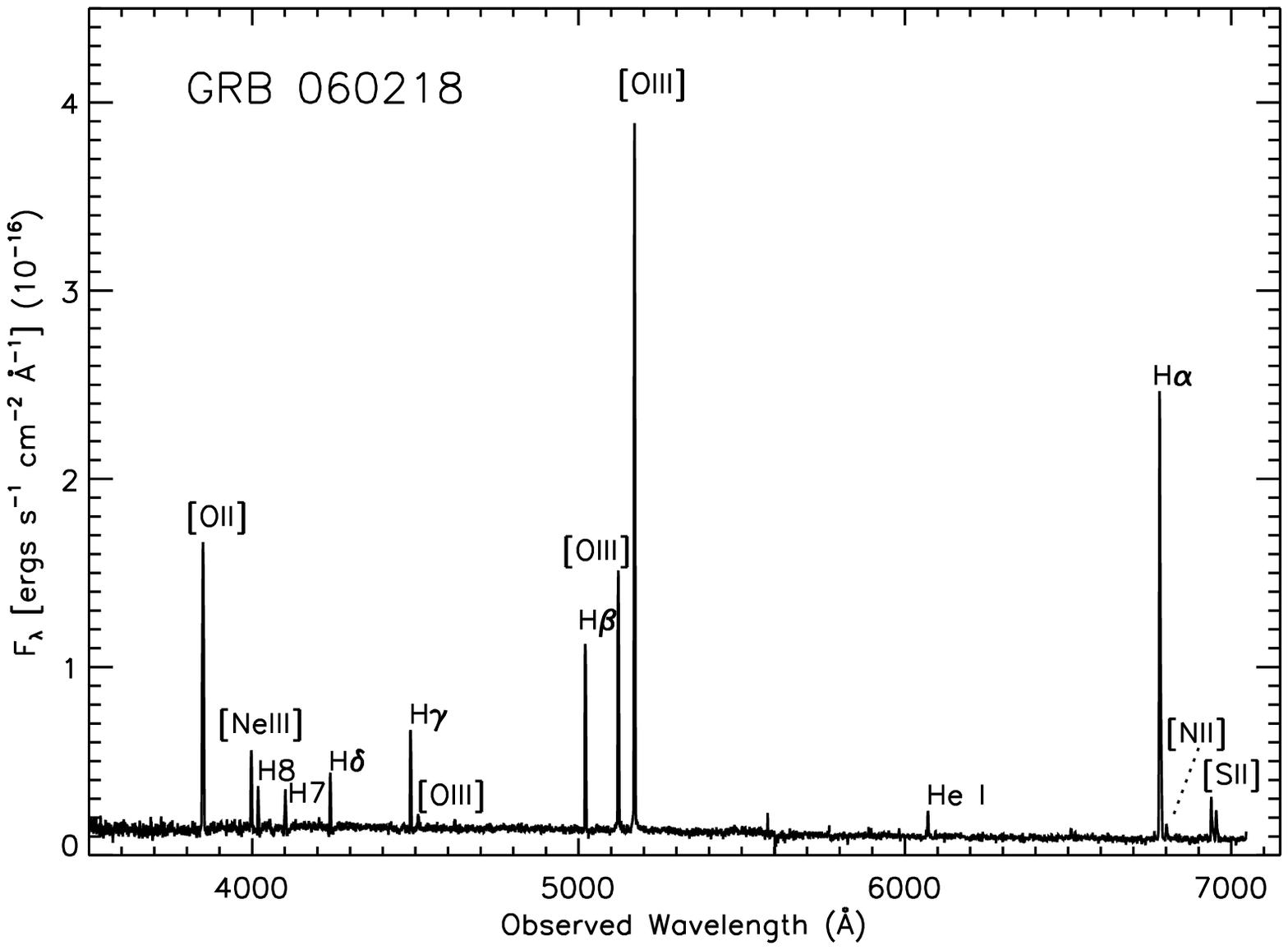}
\caption{Our spectrum of the $z = 0.034$ host galaxy of GRB 060218, observed with LRIS at Keck on 7 September 2007.}
\end{figure}


\begin{references}
\reference {} Adelman-McCarthy, J. K. et al.\ 2006, ApJS, 162, 38
\reference {} Allende Prieto, C., Lambert, D. L., \& Asplund, M. 2001, ApJ, 556, L63
\reference {} Asplund, M., Grevesse, N., \& Sauval, A. J. 2005, in Astronomical Society of the Pacific Conference Series, Vol. 336, Cosmic Abundances as Records of Stellar Evolution and Nucleosynthesis, ed. T. G. Barnes, III \& F. N. Bash, 25
\reference {} Baldwin, J. A., Phillips, M. M., \& Terlevich, R. 1981, Pub. A. S. P., 93, 5
\reference {} Berger, E. 2009, ApJ, 690, 231
\reference {} Berger, E., Cowie, L. L., Kulkarni, S. R., Frail, D. A., Aussel, H., \& Barger, A. J. 2003, ApJ, 588, 99
\reference {} Berger, E., Fox, D. B., Kulkarni, S. R., Frail, D. A., \& Djorgovski, S. G. 2007, ApJ, 660, 504
\reference {} Berger, E., Penprase, B. E., Cenko, S. B., Kulkarni, S. R., Fox, D. B., Steidel, C. C., Reddy, N. A. 2006, ApJ, 642, 979
\reference {} Bloom, J. S., Kulkarni, S. R., \& Djorgovski, S. G. 2002, ApJ, 121, 1111
\reference {} Brown, W., Kewley, L. J., \& Geller, M. J. 2008, AJ, 135, 92
\reference {} Cardelli, J. A., Clayton, G. C., \& Mathis, J. S. 1989, ApJ, 345, 245
\reference {} Castro-Tirado, A. J. et al.\ 2007, A\&A, 475, 101
\reference {} Cobb, B. E., Bailyn, C. D., van Dokkum, P. G., Buxton, M. M., \& Bloom, J. S. 2004, ApJ, 608, L93
\reference {} Chary, R., Berger, E., \& Cowie, L. 2007, ApJ 671, 272
\reference {} Chen, H-W., Prochaska, J. X., Bloom, J. S., \& Thompson, I. B. ApJ, 634, L25
\reference {} Christensen, L., Hjorth, J., Gorosabel, J. 2004, A\&A, 425, 913
\reference {} Christensen, L., Vreeswijk, P. M., Sollerman, J., Th\"{o}ne, C. C., Le Floc'h, E., \& Wiersema, K. 2008, A\&A, 490, 45
\reference {} Copetti, M. V. F., Pastoriza, M. G., \& Dottori, H. A. 1986, A\&A, 156, 111
\reference {} Dopita, M. A., Kewley, L. J., Heisler, C. A., \& Sutherland, R. S. 2000, ApJ, 542, 224
\reference {} Erb, D. K., Shapley, A. E., Pettini, M., Steidel, C. C., Reddy, N. A., \& Adelberger, K. L. 2006, ApJ, 644, 813
\reference {} Fruchter, A. S. et al.\ 2006, Nature, 441, 463
\reference {} Fynbo, J. P. U., et al.\ 2006, A\&A, 451, 47
\reference {} Fynbo, J. P. U., Hjorth, J., Malesani, D., Sollerman, J., Watson, D., Jakobsson, P., Gorosabel, J., \& Jaunsen, A. O. 2007, arXiv:astro-ph/0703458v2
\reference {} Galama, T. J. et al. 1998, Nature, 395, 670
\reference {} Garnett, D. R. 1992, AJ, 103, 1330
\reference {} Garnett, D. R., Kennicutt, R. C., \& Bresolin, F. 2004, ApJ, 607, L21
\reference {} Giavalisco, M. et al.\ 2004, ApJ, 600, 93
\reference {} Gordon, D. \& Gottesman, T. S. 1981, AJ, 86 161
\reference {} Gorosabel, J. et al.\ 2005, A\&A, 444, 711
\reference {} Graham, J. et al.\ 2009a, in Gamma-Ray Burst: Sixth Huntsville Symposium, American Institute of Physics Conference Proceedings 1133, p. 269
\reference {} Graham, J. et al.\ 2009b, in prep
\reference {} Jansen, R. A., Fabricant, D., Franx, M., \& Caldwell, N. 2000a, ApJS, 126, 331
\reference {} Jansen, R. A., Franx, M., Fabricant, D., \& Caldwell, N. 2000b, ApJS, 126, 271
\reference {} Hammer, F., Flores, H., Schaerer, D., Dessauges-Zavadsky, M., Le Floc'h, E., \& Puech, M. 2006, A\&A, 454, 103
\reference {} Hirschi, R., Meynet, G., \& Maeder, A. 2005, A\&A, 443, 581
\reference {} Huchra, J. P., Davis, M., Latham, D., \& Tonry, J. 1983, ApJS, 52, 89
\reference {} Hjorth, J. et al. 2003, Nature, 423, 847
\reference {} Kauffmann, G. et al.\ 2003, MNRAS, 346, 1055
\reference {} Kawabata, K. S. et al.\ 2003, ApJ, 593, L19
\reference {} Kawai, N., Kosugi, G., Aoki, K. et al., 2006, Nature, 440, 184
\reference {} Kennicutt, R. C. 1998, ARA\&A, 36, 189
\reference {} Kennicutt, R. C., Bresolin, F., \& Garnett, D. R. 2003, ApJ, 591, 801
\reference {} Kewley, L. J., Brown, W. R., Geller, M. J., Kenyon, S. J., \& Kurtz, M. J. 2007, AJ, 133, 882
\reference {} Kewley, L. J. \& Dopita, M. A. 2002, ApJS, 142, 35
\reference {} Kewley, L. J., Dopita, M. A., Sutherland, R. S., Heisler, C. A., \& Trevena, J. 2001, ApJ, 556, 121
\reference {} Kewley, L. J., Ellison, S. L. 2008, ApJ, 681, 1183
\reference {} Kewley, L. J., Geller, M. J., \& Jansen, R. A. 2004, AJ, 127, 2002
\reference {} Kewley, L. J., Groves, B., Kauffman, G., \& Heckman, T. 2006, MNRAS, 372, 961
\reference {} Kinney, A. L. et al.\ 1993, ApJS, 86, 5
\reference {} Kobulnicky, H. A. \& Kewley, L. J. 2004, ApJ, 617, 24
\reference {} Kocevski, D., West, A. A., \& Modjaz, M. 2009, ApJ, 702, 377
\reference {} Kouveliotou, C. et al.\ 1993 ApJ, 413, L101
\reference {} Kong, X. \& Cheng, F. Z. 2002, A\&A, 389, 845
\reference {} Kudritzki, R. P. 2002, ApJ, 557, 389
\reference {} Kudritzki, R. P. \& Puls, J. 2000, ARA\&A, 38, 613
\reference {} K\"{u}pc\"{u} Yoldas, A., Greiner, J., \& Perna, R. 2006, A\&A, 457, 115
\reference {} Kurucz, R. L. 1991, in Stellar Atmospheres: Beyond Classical Limits, ed. L. Crivellari, I. Hubeny, \& D. G. Hummer (Dordrecht: Kluwer), 441
\reference {} Langer, N. \& Norman, C. A. 2006, ApJ, 638, L63
\reference {} Leitherer, C. 2008, {\it IAU Symp. 255, Low-Metallicity Star Formation: From the First Stars to Dwarf Galaxies}, ed. L. K. Hunt, S. Madden, \& R. Schneider (Cambridge: CUP), 305
\reference {} Leitherer, C., Schaerer, D., Holdader, J. D., Delgado, R. M.G., Robert, C., Kune, D. F., de Mello, D. F., Devost, D., \& Heckman, T. M., et al.\ 1999, ApJS, 123, 3
\reference {} Leitherer, C, Carmelle, R., \& Drissen, L. 1992, ApJ, 401, 596
\reference {} Levesque, E. M. \& Kewley, L. J. 2007, ApJ, 667, L121
\reference {} Levesque, E. M., Kewley, L. J., \& Larson, K. 2009, submitted
\reference {} Levesque, E. M., Massey, P., Olsen, K. A. G., \& Plez, B. 2007, ApJ, 667, 202
\reference {} Liu, X., Shapley, A. E., Coil, A. L., Brinchmann, J., Ma, C.-P. 2008, ApJ, 678, 758
\reference {} Margutti, R. et al.\ 2007, A\&A, 474, 815
\reference {} Massey, P. 2003, ARA\&A, 41, 15
\reference {} McGaugh, S. S. 1991, ApJ, 380, 140
\reference {} Meynet, G. \& Maeder, A. 2000, A\&A, 361, 101
\reference {} Meynet, G. \& Maeder, A. 2005, A\&A, 429, 581
\reference {} Modjaz, M. et al.\ 2006, ApJ, 645, L21
\reference {} Modjaz, M., Kewley, L. J., Kirshner, R. P., Stanek, K. Z., Challis, P., Garnavich, P. M., Greene, J. E., Kelly, P. L., Prieto, J. L. 2008, AJ, 135, 1136
\reference {} Ofek, E. O. et al.\ 2007, ApJ, 662, 1129
\reference {} Osterbrock, D.1989, Astrophysics of gaseous nebulae and active galactic nuclei (University Science Books)
\reference {} Prochaska, J. X. et al.\ 2004, ApJ, 611, 200
\reference {} Prochaska, J. X., Chen, H-W., Dessauges-Zavadsky, M., Bloom, J. S. 2007, ApJ, 666, 267
\reference {} Pettini, M. \& Pagel, B. E. J. 2004, MNRAS, 348, 59
\reference {} Richer, M. G. \& McCall, M. L. 1995, ApJ, 445, 642
\reference {} Savaglio, S., Glazebrook, K., \& Le Borgne, D. 2006, in American Institute of Physics Conference Series, ed. S. S. Holt, N. Gehrels, \& J. A. Nousek, 540
\reference {} Savaglio, S., Glazebrook, K., \& Le Borgne, D. 2009, ApJ, 1091, 182
\reference {} Schaerer, D. 1996a, A\&A, 309, 129
\reference {} Schaerer, D. 1996b, in From Stars to Galaxies: The Impact of Stellar Physics on Galaxy Evolution, ASP Conf. Series 98, eds. C. Leitherer, U. Fritze - von Alvensleben, J. Huchra, p. 174
\reference {} Schaerer, D. \& de Koter, A. 1997, A\&A, 322, 598
\reference {} Schaerer, D. \& Vacca, W .D. 1998, ApJ, 497, 618
\reference {} Schlegel, D. J., Finkbeiner, D. P., \& Davis, M. 1998, ApJ, 500, 525
\reference {} Schmutz, W., Leitherer, C., \& Gruenwald, R. 1992, PASP, 104, 1164
\reference {} Shapley, A. E., Erb, D. K., Pettini, M., Steidel, C. C., \& Adelberger, K. L. 2004, ApJ, 612, 108
\reference {} Shi, F., Kong, X., \& Cheng, F. Z. 2006, A\&A, 453, 487
\reference {} Soderberg, A. et al.\ 2004, ApJ, 606, 994
\reference {} Stanek, K. Z. et al.\ 2003, ApJ, 591, L17
\reference {} Stanek, K. Z., et al.\ 2006, Acta Astron. 56, 333
\reference {} Stasinska, G. 1980, A\&A, 84, 320
\reference {} Stasinska, G. \& Leitherer, C. 1996, ApJS, 107, 661
\reference {} Thuan, T. X. et al.\ 1999, A\&AS, 139, 1
\reference {} Thuan, T. X. \& Martin, G. E. 1981, ApJ, 247, 923
\reference {} van den Heuvel, E. P. J. \& Yoon, S.-C. 2007, Ap\&SS, 311, 177
\reference {} van Zee et al.\ 1998, AJ, 116, 2805
\reference {} Veilleux, S. \& Osterbrock, D. E. 1987, ApJS, 63, 295
\reference {} Vink, J. S., de Koter, A., \& Lamers, H. J. G. L. M. 2001, A\&A, 369, 574
\reference {} Vink, J. S. \& de Koter, A. 2005, A\&A, 442, 587
\reference {} Wainwright, C., Berger, E., Penprase, B. E. 2007, ApJ, 657, 367
\reference {} Wiersema, K. et al.\ 2007, A\&A, 464, 529
\reference {} Wijers, R. A. M. J., Bloom, J. S., Bagla, J. S., \& Natarajan, P. 1998, MNRAS, 297, L13
\reference {} Wirth, G. D. et al.\ 2004, AJ, 127, 3121
\reference {} Woosley, S. E. 1993, ApJ, 405, 273
\reference {} Woosley, S. E. \& Heger, A. 2006, ApJ, 637, 914
\reference {} Yoon, S.-C., Langer, N., \& Norman, C. 2006, Astron. Ap. 460, 199

\end{references}
\end{document}